\documentclass[12pt]{article}
\usepackage[utf8]{inputenc}
\usepackage[margin=1in]{geometry}
\usepackage{multirow}
\usepackage{amssymb}
\usepackage{amsmath}
\usepackage{graphicx}

\usepackage[authoryear,round]{natbib} 
\usepackage[hyphens]{url}   
\usepackage{xurl} 
\usepackage{verbatim}
\usepackage{lscape}
\usepackage{array}
\usepackage{float}       
\usepackage{placeins}    
\usepackage{lipsum}      
\usepackage{setspace}
\usepackage{titlesec}
\usepackage{booktabs}
\usepackage{soul}
\usepackage{xcolor}
\sethlcolor{yellow}
\usepackage{rotating}
\usepackage{longtable}
\usepackage[mode=tex]{standalone}   
\usepackage{tikz}\usetikzlibrary{arrows.meta, positioning, shapes.geometric,fit,calc,intersections}  
\usepackage[hidelinks]{hyperref}
\usepackage{xr-hyper} 
\usepackage[nameinlink,noabbrev]{cleveref}

\titlelabel{\thetitle.\quad}

\tikzset{
  every node/.style={font=\sffamily},
  startstop/.style = {rectangle,rounded corners,minimum width=1cm,minimum height=0.4cm,
                      text centered,draw=black,fill=blue!20,align=center,inner sep=5pt},
  io/.style        = {trapezium,trapezium left angle=70,trapezium right angle=110,
                      minimum width=1cm,minimum height=0.4cm,text centered,draw=black,
                      fill=blue!30,align=center},
  process/.style   = {rectangle,minimum width=1cm,minimum height=0.4cm,text centered,
                      draw=black,fill=orange!30,align=center,inner sep=5pt},
  process1/.style  = {rectangle,minimum width=1cm,minimum height=0.4cm,text centered,
                      draw=black,fill=green!30,align=center},
  coord/.style     = {circle,minimum size=0.5mm,inner sep=0mm,fill=black,align=center},
  annotate1/.style = {rectangle,minimum width=1cm,minimum height=0.4cm,draw=white,
                      fill=white,align=center},
  decision/.style  = {diamond,text centered,draw=black,fill=green!30,aspect=3,
                      align=center,inner sep=0.1ex},
  annotate/.style  = {rectangle,minimum width=1cm,minimum height=0.4cm,draw=black,
                      fill=white,align=center},
  arrow/.style     = {thick,->,>=Stealth}
}

\pgfkeys{
  /split ellipse/.is family, /split ellipse,
  rx/.initial=3cm,                  
  ry/.initial=2cm,                  
  gap/.initial=8pt,                 
  left color/.initial=orange!35,
  right color/.initial=red!25,
  outline/.initial=1.2pt,           
  label font/.initial=\fontsize{32}{32}\selectfont\bfseries,
  left label/.initial=A,
  right label/.initial=B,
}

\tikzset{
  split ellipse/.pic={
    \def\ax{\pgfkeysvalueof{/split ellipse/rx}}
    \def\by{\pgfkeysvalueof{/split ellipse/ry}}
    \def\byy{\pgfkeysvalueof{/split ellipse/rx}}
    \coordinate (-west)  at (-\byy,0);
    \coordinate (-east)  at (\byy,0);
    \coordinate (-center) at (0,0);
    \coordinate (-south)  at (0,-\by);
    \coordinate (-north)  at (0,\by);
    \def\g {\pgfkeysvalueof{/split ellipse/gap}}
    \def\lc{\pgfkeysvalueof{/split ellipse/left color}}
    \def\rc{\pgfkeysvalueof{/split ellipse/right color}}
    \def\ow{\pgfkeysvalueof{/split ellipse/outline}}
    \def\labfont{\pgfkeysvalueof{/split ellipse/label font}}
    \def\L{\pgfkeysvalueof{/split ellipse/left label}}
    \def\R{\pgfkeysvalueof{/split ellipse/right label}}

    \pgfmathsetmacro{\rxpt}{\ax/1pt}
    \pgfmathsetmacro{\gpt}{\g/1pt}
    \pgfmathsetmacro{\ratio}{min(.9999, max(-.9999, \gpt/\rxpt))}
    \pgfmathsetlengthmacro{\yg}{\by*sqrt(1 - (\ratio)^2)}
    \pgfmathsetmacro{\alpha}{acos(\ratio)}               
    \pgfmathsetmacro{\tLu}{180 - \alpha}                 
    \pgfmathsetmacro{\tLl}{180 + \alpha}                 
    \pgfmathsetmacro{\tRu}{\alpha}                       
    \pgfmathsetmacro{\tRl}{360 - \alpha}                 

    \path[fill=\lc, draw=black, line width=\ow, line join=round, line cap=round]
      (-\g,-\yg) -- (-\g,\yg)
      plot[samples=120, smooth, variable=\t, domain=\tLu:\tLl]
        ({\ax*cos(\t)}, {\by*sin(\t)})
      -- cycle;

    \path[fill=\rc, draw=black, line width=\ow, line join=round, line cap=round]
      (\g,-\yg) -- (\g,\yg)
      plot[samples=120, smooth, variable=\t, domain=\tRl:360+\tRu]
        ({\ax*cos(\t)}, {\by*sin(\t)})
      -- cycle;

    \node[font=\labfont] at (-0.5*\ax, 0) {\L};
    \node[font=\labfont] at ( 0.5*\ax, 0) {\R};
  }
}

\pgfkeys{/split ellipse/solid fill color/.initial=}

\tikzset{
  solid ellipse/.pic={
    \def\ax{\pgfkeysvalueof{/split ellipse/rx}}
    \def\by{\pgfkeysvalueof{/split ellipse/ry}}
    \def\byy{\pgfkeysvalueof{/split ellipse/rx}}
    \coordinate (-east)  at (\byy,0);
    \coordinate (-west)  at (-\byy,0);
    \coordinate (-center) at (0,0);
    \coordinate (-south)  at (0,-\by);
    \coordinate (-north)  at (0,\by);   
    \def\ow{\pgfkeysvalueof{/split ellipse/outline}}
    \def\labfont{\pgfkeysvalueof{/split ellipse/label font}}
    \def\L{\pgfkeysvalueof{/split ellipse/left label}}
    \def\R{\pgfkeysvalueof{/split ellipse/right label}}
    \def\lc{\pgfkeysvalueof{/split ellipse/left color}}

    \edef\maybeSolid{\pgfkeysvalueof{/split ellipse/solid fill color}}
    \ifx\maybeSolid\empty
      \def\fillcol{\lc}
    \else
      \def\fillcol{\maybeSolid}
    \fi

    \path[fill=\fillcol, draw=black, line width=\ow, line join=round, line cap=round]
      (0,0) ellipse [x radius=\ax, y radius=\by];

    \node[font=\labfont] at (-0.5*\ax, 0) {\L};
    \node[font=\labfont] at ( 0.5*\ax, 0) {\R};
  }
}

\begin{document}

\begin{center}
{\large \bf  Adaptively Combining Randomized and External Control Data Using a Mixture Prior in the Presence of Heterogeneity}
\end{center}

 \begin{center}
 Xinyi He$^{1,2}$, Peter F. Thall$^2$, Ying Yuan$^{2}$, Suyu Liu$^{2}$

 {\em
 $^1$Graduate School of Biomedical Sciences, The University of Texas Health Science Center at Houston\\
 Houston, TX 77030, USA \\

 $^2$Department of Biostatistics, The University of Texas MD Anderson Cancer Center\\
 Houston, TX 77030, USA \\
}
\end{center}

\doublespacing

\begin{center}
\textbf{Abstract}
\end{center}

\vspace{1mm}

When resources are limited in a randomized trial, a common strategy is to augment data from the control arm with external controls.  
We propose a hybrid  group sequential Bayesian design for comparing time-to-event distributions
that uses this approach. 
The design accommodates settings where between-treatment effects 
may differ between patient subgroups, and adaptively combines subgroups having similar 
hazard functions.
The aim is to reduce the  control arm sample size without compromising the validity of comparative inferences
due to systematic trial-versus-external data differences. To dynamically borrow external control data, 
the model uses
a self-adapting mixture prior, defined as a weighted average of a non-informative component and an informative component constructed from the external controls.
The amount of borrowing is proportional to the agreement between the randomized and external controls. 
 At each interim analysis, the design compares the experimental and control treatments in each subgroup,
and if inferiority or superiority is concluded stops accrual early for that subgroup.
Simulations show that the proposed design outperforms other methods for incorporating external control data in a randomized trial,
and it improves efficiency.

\vspace{5mm}

\clearpage
\section{Introduction}
Randomization is the established basis for making causally valid  comparisons 
in clinical trials
(\cite{Rubin1978, Armitage2003}).  For a randomized controlled trial (RCT) to identify a small to moderate difference between  an experimental treatment $E$ and a control $C$ with high power,  a large sample size is required.
Such trials may be prohibitively expensive or time consuming.   This arises
in studies of rare diseases or   small subpopulations. 
 This problem may be addressed by combining external controls and randomized controls, which  hybridizes a conventional RCT and a one-arm study to compare $E$ to an external control.
 
In recent years, hybrid RCTs have gained acceptance by the pharmaceutical industry and regulatory agencies (\cite{jahanshahi2021use,schmidli2020beyond}). The U.S. Food and Drug Administration (FDA), under its Real-World Evidence program, has issued guidelines for using data  from electronic health records and medical claims for regulatory decision-making about drugs and biologics (\cite{fda2021guidance,fda2023guidance}). They have established standards for submission of study data derived from real world data sources (\cite{fda2023guidance-2}).

A key methodological challenge in hybrid RCTs is accounting for 
systematic  differences between the trial control arm and external control data that may
lead to biased  between-treatment effect estimates. 
A variety of Bayesian information-borrowing methods have been proposed to address this issue. The power prior and its variants incorporate the likelihood of external data raised to a power parameter,  as a weighting factor to account for disagreement between trial and external controls (\cite{ibrahim2000power}).
\cite{Hobbs2011} introduced commensurate priors, which determine the amount of information borrowed from external data using a shrinkage parameter. 
\cite{murray2014semiparametric} extended this to time-to-event endpoints, assuming a piecewise exponential (PE) model. \cite{jiang2023elastic} proposed elastic priors,  which include a monotone function of the degree  of agreement between trial and external  controls that determines the degree of borrowing.
Elastic priors have been used, for example, in biosimilar clinical trial designs (\cite{zhang2021elastic,zhang2022bayesian}).

Mixture priors take an empirical Bayes approach to address prior-versus-data conflicts when incorporating external controls. A mixture prior is  a weighted average of a noninformative prior and an informative prior derived from the external controls. 
 \cite{Schmidli2014} introduced the robust meta-analytic predictive (MAP) prior, which is a mixture of a vague prior for the control effect and a MAP prior based on external data, using prespecified mixture weights for the two components. This approach has been applied, for example, to design a pediatric multiple sclerosis trial (\cite{schmidli2020beyond, fda2023pediatricms}). A key challenge of the robust MAP prior is the need to specify the mixture weights.  
 To address this, \cite{yang2023SAM} proposed self-adapting mixture (SAM) priors, which employ data-driven mixture weights derived from likelihood ratio test statistics or Bayes factors, allowing the degree of borrowing to be adjusted according to the level of trial versus external control data conflict. \cite{chiaruttini2025bayesian} used  this idea to develop a Bayesian group sequential design incorporating the SAM prior framework for medical device trials. Applications accommodating time-to-event endpoints have not been explored, however.

Many frequentist methods for incorporating external control data  have been developed. \cite{viele2014use}, \cite{li2020revisit}, \cite{ventz2022design}, and \cite{yang2023elastic} used  tests to assess similarity between external and trial controls, and decide  how to incorporate external data. \cite{li2022conditional} 
proposed a conditional borrowing approach, with weights of external and trial controls determined by the similarity between empirical distributions of log hazard ratio estimates derived through resampling. \cite{stuart2008matching}, \cite{yuan2019design}, \cite{chen2020propensity}, and \cite{guo2024adaptive} used propensity score matching between trial and external control patients  to correct for bias in treatment effect estimates. \cite{zhang2025outcome} discussed  robust regression models in the context of the G-computation method (\cite{robins1986new}), and a doubly robust weighted regression method (\cite{hirano2001estimation}) for hybrid control estimation.

In this paper,  we propose a Bayesian group sequential hybrid control design for RCTs with time-to-event outcomes. We are
motivated by the Block Heart Failure trial (\cite{Curtis2013}), described in Section \ref{motivatingtrial}. Our design borrows external control data dynamically using a SAM prior
for the parameters of  a time-to-event likelihood. This prior uses the estimated degree of prior-versus-data disagreement to  adjust weights for a non-informative prior and an informative prior derived from external controls, with the external data prior given a weight proportional to the amount of agreement.   To ensure robustness, the SAM prior accommodates non-constant hazards. We assume that the patient population has been partitioned into subgroups, possibly  using prognostic variables, biomarkers, or disease subtypes. 
The design accommodates  heterogeneity of $E$ or $C$ effects on failure time between subgroups, and includes subgroup-specific rules for adaptive interim stopping if superiority or inferiority of $E$ compared to $C$ is concluded.   The design adaptively clusters similar subgroups, depending on  whether the outcome distributions of  $E$,  $C$, or  $E$-versus-$C$ effects  differ between  subgroups.  Clustering  borrows strength between subgroups in the same cluster
and reduces the model dimension.

The remainder of the paper is organized as follows. In Section \ref{motivatingtrial}, we describe the motivating trial. The statistical model and SAM prior are described in Section \ref{model}. The design is presented in Section \ref{trialdesign}. Results of a simulation study to evaluate the design's operating characteristics (OCs) under a range of scenarios in application are summarized in Section \ref{simulationstudy}. We close with a brief discussion in Section \ref{s:discuss}.

\section{Motivating Trial} \label{motivatingtrial}
The Block Heart Failure (HF) trial (\cite{Curtis2013}) investigated benefits of biventricular (BiV) pacing, also known as cardiac resynchronization therapy (CRT),  compared to traditional right ventricular (RV) pacing in patients with previous heart failure and atrioventricular (AV) blockage. The  aim  was to determine whether BiV pacing could improve outcomes, compared to RV pacing, by more effectively coordinating the heart's contractions, thereby enhancing cardiac function and reducing morbidity and mortality. Patients were randomized to receive either $C$ = RV pacing  or $E$ = BiV pacing.  
The primary outcome was  time to first treatment failure, defined as death from any cause, an increase in the left ventricular endsystolic volume of 15\% or more, or an urgent care visit for heart failure requiring intravenous therapy.
Patients were classified  as being eligible for being implanted with either (1)  a CRT pacemaker (CRT-P) or (2) a CRT defibrillator (CRT-D), with 
the BiV versus RV treatment comparison  made within each subgroup, and an overall average mean effect also estimated.  These subgroups are disjoint because  a patient is classified as CRT-D if they have an indication for defibrillation therapy. 
The Block HF trial design was based on a piecewise exponential (PE) model  to account for non-constant hazards of the failure time distributions  in the $E$ and $C$ arms. It was assumed that the BiV-versus-RV ($E$-versus-$C$) treatment effect followed a proportional hazards (PH) model, with the same log hazard ratios $\theta_{CRT-P}$ and $\theta_{CRT-D}$ for the two subgroups in all time intervals, and an average effect, $\theta$. Up to two interim analyses were planned, with maximum sample size 1200 and follow-up 12 months. The primary comparison was based on the marginal posterior distribution of $\theta$. 

Sample size calculations for the Block HF trial were based on event rates from the historical trials MIRACLE (\cite{Abraham2002}) and MIRACLE-ICD (\cite{young2003}). The MIRACLE trial was a randomized, double-blind, placebo-controlled study comparing CRT ($E$ = intervention) to a standard pacemaker ($C$ = control). The MIRACLE-ICD trial (\cite{young2003}), assessed the combination of CRT and an Implantable Cardioverter-Defibrillator (ICD) in heart failure patients at risk of sudden cardiac death. These  studies were selected because they involved CRT therapy in NYHA Class III and IV patients, and  provided data on mortality, heart failure-related urgent care, and LVESVI, which aligned with the primary outcomes of the Block HF trial. The MIRACLE trial informed the CRT-P subgroup, and the MIRACLE-ICD trial informed the CRT-D subgroup.

Motivated by the Block HF trial, the objective of the design proposed in this paper is to incorporate the historical data from the MIRACLE and MIRACLE-ICD trials to construct an adaptive hybrid control trial design.  The design  includes subgroup-specific early stopping 
for superiority or inferiority, and it accounts for whether $C$ or $E$ effects on failure time  are homogeneous or heterogeneous across subgroups.

\section{Statistical Model}
\label{model}
\subsection{Event Time Likelihood}
For each patient $i=1,\dots,n$, let $g_i\in\{1,\dots,G\}$ denote their subgroup,
and let  $k_i \in \{E, C\}$ their assigned treatment. Let $Z_i$ denote their observed time to failure or censoring, with failure event indicator $d_i=1$ if $Z_i$  is observed and $d_i=0$ if it is censored. Let $n_k=\sum_{i=i}^n\mathbb{I}(k_i=k)$ be the number of patients in arm $k$. 
Given subgroups $\{g_i\}$, 
there amy be uncertainty whether treatment effects are homogeneous or heterogeneous across  subgroups.
If subgroups have homogeneous treatment effects, pooling their data  can substantially improve the efficiency of treatment effect estimates. To allow this possibility and facilitate adaptive information sharing between subgroups, we introduce latent cluster membership variables, so that subgroups within the same cluster share a common treatment effect in each treatment arm.
We first present the model assuming  that the cluster memberships are known, and later describe methods for adaptively determining  cluster memberships.

For each subgroup $g$ in arm $k$, we define a cluster membership variable $s_{g,k}\in\{1,\dots,G\}$, and denote ${\boldsymbol{s}}_{k}=(s_{1,k},\dots,s_{G,k})$. If $s_{g,k}=s_{g',k}$ for $g\neq g'$, subgroups $g$ and $g'$ have the same event time distribution and  are combined into one arm $k$ cluster, indexed by $g'$.  We use the smaller subgroup index $g' < g$ to identify the cluster uniquely.
If $s_{1,k}=\dots=s_{G,k}=1$, this corresponds to complete homogeneity of the arm $k$ hazards. If $s_{g,k}=g$ for all $g\in\{1,\dots,G\}$,  the $G$ subgroups have  different arm $k$ event time distributions. 
Let $\mathcal{G}_{k}=\{g: s_{g,k}=g\}$ denote the set of induced cluster indices in arm $k$.  The elements of $\mathcal{G}_k$ serve as cluster indices, and $|\mathcal{G}_k|\leq G$  is its cardinality. We define $\mathcal{G}_k$ because it includes the unique cluster identifying indices in arm $k$. In particular, $\mathcal{G}_k$ always includes the subgroup index $g=1$.
For  $G=2$ and $k=C$, if $\mathcal{G}_C=\{1,2\}$ then subgroups 1 and 2 have different event time distributions in $C$ and  belong to distinct clusters with indices 1 and 2. If $\mathcal{G}_{C}=\{1\}$,  subgroups 1 and 2 have the same event time distribution in $C$ and are grouped into a cluster indexed by 1. 

The priors and likelihoods of arm $k$ depend on $\mathcal{G}_k$, which depends on $\boldsymbol{s}_k$.
To obtain a robust design, we assume failure time follows a PE distribution.
For a fixed positive integer $J$, we partition $[0,~\infty)$ into $J$ sub-intervals: $I_1=[t_0,t_1),~I_2=[t_1,t_2),\dots,I_J=[t_{J-1},t_J)$, where $t_0=0$ and $t_J=\infty$. 
Using $s=1$ as the reference cluster, we denote the hazard ratio of cluster $s > 1$ versus cluster 1 in arm $k$ by $\exp{(\theta_{s,k}})$, and set $\theta_{1,k}=0$.
We assume the hazard function for cluster $s\in\mathcal{G}_k$ follows a PH model,
$h_{k,s}(t)=\lambda_{j,k}\exp(\theta_{s,k}),$
{\rm for}
$t\in I_j,$
where $\lambda_{j,k}$ are the baseline hazards. 
Let  $d_{i,j}=1$ if the $i$th patient has the event in  $I_j$, and $d_{i,j}=0$ otherwise. Let $u_{i,j}$ denote the observed time in $I_j$ for patient $i$, defined by
\begin{align}
u_{i,j}=\begin{cases}
    t_j-t_{j-1}~ &{\text{if}}~Z_{i}\geq t_j,~j=1,\dots,J-1\\
    Z_{i}-t_{j-1}~&{\text{if}}~Z_{i}\in I_j,~j=1,\dots,J\\
    0~&\text{otherwise}.
    \end{cases}
\end{align}

Assuming there are $\geq 2$ induced clusters in arm $k$,  let $\boldsymbol{\lambda}_{k}=(\lambda_{1,{k}},\dots,\lambda_{J,{k}})$ be the baseline hazard vector and {let $\boldsymbol{\theta}_k$ denote the vector of $\theta_{s,k}$’s, excluding $\theta_{1,k}$ = 0, writing $\boldsymbol{\theta}_k=(\theta_{2,k},\dots,\theta_{|\mathcal{G}_k|,k})$ for convenience with a slight abuse of notation.} Denote the failure time data of patients from arm $k$ in  $I_j$ by ${D_{j,k}}=\{(u_{i,j},d_{i,j}):~k_i=k\}$, and $\mathcal{D}_{k}=\{{D_{1,k}},\dots,{D_{J,k}}\}$. The likelihood for arm $k$ under $\mathcal{G}_k$ and $\boldsymbol{s}_k$  is 
\begin{align}
    {\cal L}(\boldsymbol{\lambda}_{k},\boldsymbol{\theta}_k\mid \mathcal{D}_k,{\boldsymbol{s}_k}) = \prod_{{s}=2}^{|\mathcal{G}_k|}\prod_{i\in V_{s,k}}\prod_{j=1}^J(\lambda_{j,k}e^{\theta_{s,k}})^{d_{i,j}}\exp(-\lambda_{j,k}e^{\theta_{{s},k}}u_{i,j}),\label{eq:likelihood}
\end{align}
for $V_{s,k}=\{i:k_i=k,s_{g_i,k}= s\}$ the patient indices with cluster index $s$ in arm $k$, 
and $\theta_{1,k}\equiv0$,  
with full likelihood  
$\prod_{k=E,C}
 {\cal L}(\boldsymbol{\lambda}_k,\boldsymbol{\theta}_{{k}}\mid\mathcal{D}_{k},{\boldsymbol{s}_k}).$

\subsection{SAM prior to incorporate historical control data}
We incorporate historical control data $\mathcal{D}^{hist}$ by assuming a SAM prior for  $\boldsymbol{\lambda}_C$ and $\boldsymbol{\theta}_{{C}}$,  
defined as a mixture of an informative  and a noninformative prior. The noninformative component uses conventional priors, 
\[
\pi_0(\lambda_{j,C}) =  {\rm{Gamma}}(\alpha_0,\beta_0), \ j=1,\dots,J \  \ {\rm and} \  \
\pi_0(\theta_{s,C}) = \mathcal{N}(\mu_0,\sigma_0^2), \ {s}=2,\dots,|\mathcal{G}_{C}|,
\]
where  $\alpha_0$ and $\beta_0$ are small (e.g., 0.01), $\mu_0=0$, and $\sigma_0^2$ is large value (e.g., $10^4$). 

Extending the SAM framework to a PH model with a PE baseline hazard is nontrivial, and  requires substantial methodological development. 
We will describe how to construct the informative prior component and derive the data-dependent adaptive weight that governs the degree of borrowing from historical control data.  
The informative prior is constructed based on $\mathcal{D}^{hist}$ as follows. For simplicity, we assume  there are $|\mathcal{G}_C|\geq 2$ clusters and every control cluster has historical data. 
For  control clusters without historical data, one can use conventional noninformative priors in place of the SAM prior. We assume that the elements of $\boldsymbol{\lambda}_C$ = $(\lambda_{1,C},\dots,\lambda_{J,C})$ and {$\boldsymbol{\theta}_C=(\theta_{2,C},\dots,\theta_{|\mathcal{G}_{C}|,C})$} are independent {\it a priori}, and temporarily  simplify notation by using $\boldsymbol{\lambda}$ and $\boldsymbol{\theta}$ in place of $\boldsymbol{\lambda}_C$ and $\boldsymbol{\theta}_C$, and writing
 $\pi(\boldsymbol{\lambda},\boldsymbol{\theta}\mid \mathcal{D}^{hist},{\boldsymbol{s}_C})$ as $\pi_1(\boldsymbol{\lambda},\boldsymbol{\theta}\mid {\boldsymbol{s}_C})$.
We construct the informative prior $\pi_1(\boldsymbol{\lambda},\boldsymbol{\theta}\mid {\boldsymbol{s}_C})$  taking an empirical Bayes approach, estimating its hyperparameters from $\mathcal{D}^{hist}$ given the cluster assignments {$\boldsymbol{s}_C$}.  If $\mathcal{D}^{hist}$ is from one historical study, starting with a vague baseline prior $\pi_0(\boldsymbol{\lambda},\boldsymbol{\theta})$ one can apply Bayes' rule to obtain the data-derived informative prior,
\begin{align*}
\pi_1(\boldsymbol{\lambda},\boldsymbol{\theta}\mid {\boldsymbol{s}_C})\ \  &\propto \ \ \pi_0(\boldsymbol{\lambda},\boldsymbol{\theta}\mid {\boldsymbol{s}_C})L(\boldsymbol{\lambda},\boldsymbol{\theta}\mid\mathcal{D}^{hist},{\boldsymbol{s}_C}) \\
&= \prod_{j=1}^J\pi_0(\lambda_j)\cdot\prod_{{ s}=2}^{|\mathcal{G}_C|}\pi_0(\theta_{s})L(\boldsymbol{\lambda},\boldsymbol{\theta}\mid\mathcal{D}^{hist},\boldsymbol{s}_C).
\end{align*}
The likelihood  ${\cal L}(\boldsymbol{\lambda},\boldsymbol{\theta} \mid \mathcal{D}^{hist},{\boldsymbol{s}_C})$ is assumed to follow the same PH model with PE baseline hazard. 
If historical data come from multiple studies, one can fit a hierarchical  model to $\mathcal{D}^{hist}$ to obtain the informative component (\cite{Schmidli2014}). 
Other methods for constructing an informative prior based on historical data are given by \cite{spiegelhalter2004bayesian}, \cite{thall2012} and \cite{Falconer2022}.

An informative prior derived as above typically does not have a closed form if there are two or more  clusters based on historical data. 
In such cases, to compute informative marginal  priors, one may draw samples of $\boldsymbol{\lambda}$ and $\boldsymbol{\theta}$ from 
$\pi_1(\boldsymbol{\lambda},\boldsymbol{\theta}\mid {\boldsymbol{s}_C})$ and approximate the prior by assuming $\it{a~posteriori}$ independence between parameters,  and matching the first and second moments with those of tractable parametric distributions. We  approximate the marginal distributions of $\lambda_{1},\dots,\lambda_{J}$ using Gamma distributions, and those of $\theta_{2},\dots,\theta_{|\mathcal{G}_C|}$ using Normal distributions. This gives the approximate informative prior 
\[
\pi_1(\boldsymbol{\lambda},\boldsymbol{\theta}\mid {\boldsymbol{s}_C})=\prod_{j=1}^J{\pi_{1}(\lambda_j)}\cdot\prod_{{ s}=2}^{|\mathcal{G}_{C}|}\pi_1(\theta_{s}),
\]
where 
$\pi_{1}(\lambda_j)={\rm{Gamma}}(\alpha_j,\beta_j)$ for $j=1,\dots,J$  and 
$\pi_1(\theta_{s})$ = $\mathcal{N}(\mu_{ s},\sigma^2_{s}),$ for ${s}=2,\dots,|\mathcal{G}_{C}|.$
If desired, one may further adjust $\alpha_j,\beta_j$ and {${\sigma^2_{s}}$ to limit the borrowed sample size.

Given the informative prior $\pi_1(\boldsymbol{\lambda},\boldsymbol{\theta}\mid {\boldsymbol{s}_C})$ obtained from  ${\cal D}^{hist}$, and  non-informative priors 
$\pi_0(\lambda_j)$ and {$\pi_0(\theta_{s})$}, 
we define the self-adapting mixture (SAM) prior  
{
\[
\pi_{SAM}(\boldsymbol{\lambda},\boldsymbol{\theta} \mid {\boldsymbol{s}_C}) = \pi_{SAM}(\boldsymbol{\lambda}\mid{\boldsymbol{s}_C})\times \pi_{SAM}(\boldsymbol{\theta}\mid {\boldsymbol{s}_C})
 =  \prod_{j=1}^J\pi_{SAM}(\lambda_j) \times \prod_{s=2}^{|\mathcal{G}_{C}|} \pi_{SAM}(\theta_{s}), 
\]}
where the components are 
{
\begin{equation}
\begin{aligned}
    \pi_{SAM}({\lambda_j})=\widetilde{w}_{\lambda,j}\pi_1({\lambda_j})+(1-\widetilde{w}_{\lambda,j})\pi_0({\lambda_j}), \ \ {\rm for}\ \  j=1,\dots,J,\\
    \pi_{SAM}(\theta_{s})=\widetilde{w}_{\theta,{s}}\pi_1(\theta_{s})+(1-\widetilde{w}_{\theta,{s}})\pi_0({\theta_{s}}),~ \ \ {\rm for }\ \ {s}=2,\dots,|\mathcal{G}_{C}|.
\end{aligned}
\end{equation}}
The weights $\widetilde{w}_{\lambda,j}$'s and $\widetilde{w}_{\theta,{s}}$'s control the degree of borrowing from  ${\cal D}^{hist}$.  
If the weights are pre-specified fixed values,  this definition is equivalent to the fixed-weight mixture prior proposed by \cite{Schmidli2014}.  If each weight is set to 0, then $\pi_{SAM}$  = $\pi_0$, the noninformative prior.  
The weights $\widetilde{w}_{\lambda,j}$
or $\widetilde{w}_{\theta,s}$  approach 1 when the historical and current trial control data agree, 
and  approach 0 when the historical and trial controls are  different. 

{For simplicity, we temporarily omit the conditioning term ${\boldsymbol{s}_C}$.}
To derive the $\widetilde{w}_{\lambda,j}$'s, we specify two sets of hypotheses
for deciding whether the historical controls and trial arm $C$ controls follow the same distribution in each interval $I_j$. These are  nulls  
 $\{H_{0,\lambda,j},~j=1,\dots,J\}$ and alternatives $\{H_{1,\lambda,j},~j=1,\dots,J\}$, where each  $H_{1,\lambda,j}$ has two points, 
as follows:
\begin{align}
    H_{0,\lambda,j}:\lambda_j=\lambda_{j}^{hist}\quad {\rm and } \quad ~H_{1,\lambda,j}:\lambda_j= e^{\delta}\lambda_{j}^{hist}~{\rm{or}}~\lambda_j= e^{-\delta}\lambda_{j}^{hist} \quad {\rm for} \quad j=1,\dots,J. \label{eq:hypothesislambda}
\end{align}
We call $\delta$  a {\it clinically significant difference} (CSD) 
 between trial and historical data treatment effects in the sense  that, if either ${\lambda}_{j}\geq e^{\delta}{\lambda_{j}^{hist}}~\text{or}~{\lambda}_{j}\leq
e^{-\delta}{\lambda_{j}^{hist}}$, then it is inappropriate to borrow information from $\mathcal{D}^{hist}$ for $\lambda_{j}$ since  ${\lambda_{j}^{hist}}$
is unacceptably different from $\lambda_j$, based on the trial data for $I_j$. 
In practice, $e^{\delta}$ may be elicited from the physicians planning the trial 
as a hazard ratio  that they  consider clinically meaningful. 

Under $H_{0,\lambda,j}$, the informative prior $\pi_1({\lambda_j})$ 
based on ${\cal D}^{hist}$ agrees with the distribution of $\lambda_j$ based on
$\mathcal{D}_{C}$ with no prior-versus-data conflict, and  it is appropriate to use $\pi_1({\lambda_j})$ to borrow information. Under $H_{1,\lambda,j}$, the distribution of $\lambda_j$ based on the current trial data and $\pi_1({\lambda_j})$ differ to such a large degree that it is not  appropriate to borrow information, and one should use the non-informative prior $\pi_0({\lambda_j})$ instead. Given $\mathcal{D}_{C}$ and $\lambda_{j}^{hist}$, the amount of information borrowed from $\mathcal{D}^{hist}$ for $\lambda_{j}$ can be quantified by the relative likelihood of $H_{0,\lambda,j}$ and $H_{1,\lambda,j}$ being correct, given by the likelihood ratio (LR) statistic,
\begin{align*}
LR_{\lambda,j}
&=
\frac{
    \max\limits_{\boldsymbol{\lambda}_{-j},\,\boldsymbol{\theta}}
    p(\mathcal{D}_C \mid H_{0,\lambda,j})
}{
    \max\limits_{\boldsymbol{\lambda}_{-j},\,\boldsymbol{\theta}}
    p(\mathcal{D}_C \mid H_{1,\lambda,j})
}
\\[6pt]
&=
\frac{
    \max\limits_{\boldsymbol{\lambda}_{-j},\,\boldsymbol{\theta}}
    p(\mathcal{D}_C \mid
      \lambda_j=\lambda_j^{hist},\,
      \boldsymbol{\lambda}_{-j},\,\boldsymbol{\theta})
}{
    \max\left\{
    \begin{aligned}
    &\max\limits_{\boldsymbol{\lambda}_{-j},\,\boldsymbol{\theta}}
    p(\mathcal{D}_C \mid
      \lambda_j=e^\delta\lambda_j^{hist},\,
      \boldsymbol{\lambda}_{-j},\,\boldsymbol{\theta}),
    \\
    &\max\limits_{\boldsymbol{\lambda}_{-j},\,\boldsymbol{\theta}}
    p(\mathcal{D}_C \mid
      \lambda_j=e^{-\delta}\lambda_j^{hist},\,
      \boldsymbol{\lambda}_{-j},\,\boldsymbol{\theta})
    \end{aligned}
    \right\}
},
\end{align*}
 denoting {$\boldsymbol{\lambda}_{-j} =
(\lambda_{1},\dots,\lambda_{j-1},\lambda_{j+1},\dots,\lambda_{J}).$} To compute 
$LR_{\lambda,j}$,  an estimate of $\lambda_j^{hist}$  is needed, such as the posterior mean $\hat{\lambda}_{j}^{hist}=E\{\pi_1(\lambda_j)\}=\frac{\alpha_j}{\beta_j}$.  The denominator  of  $LR_{\lambda,j}$
accommodates the fact that the alternative hypothesis for each $j$ in expression (\ref{eq:hypothesislambda})
has two points by  providing  less aggressive information borrowing in order to control potential bias.

Similarly, we specify two sets of hypotheses for each $\theta_{s}$, $\{H_{0,\theta,s},\;s=2,\dots,|\mathcal{G}_C|\}$ and $\{H_{1,\theta,s},\;s=2,\dots,|\mathcal{G}_C|\}$, where each alternative $H_{1,\theta,s}$  has two points:
\begin{align}
    H_{0,\theta,{s}}:\theta_{{s}}=\theta_s^{hist},\quad {\rm and } \quad ~H_{1,\theta,{s}}:\theta_{{s}}=\theta_s^{hist}+\delta~{\rm{or}}~\theta_{s}=\theta_{s}^{hist}-\delta, \quad s=2,\dots,|\mathcal{G}_{C}|.
\end{align}
Since the $\theta_{s}$'s are defined on the domain  of  the ln($\lambda_{j}$)'s, 
if $\theta_{s}\geq\theta_{s}^{hist}+\delta$ or $\theta_{s}\leq\theta_{s}^{hist}-\delta$, this exceeds the CSD  and thus it is inappropriate to borrow information from $\mathcal{D}^{hist}$ for $\theta_{s}$. Denoting 
$\boldsymbol{\theta}_{-s}=
(\theta_{2},\dots,\theta_{s-1},\theta_{s+1},\dots,\theta_{|\mathcal{G}_{C}|})$, 
the likelihood ratios are defined 
as 
\begin{align*}
    LR_{\theta,{s}}
    &=\frac{\max\limits_{\boldsymbol{\lambda},\boldsymbol{\theta}_{-{s}}}
            p(\mathcal{D}_C \mid H_{0,\theta,{s}})}
           {\max\limits_{\boldsymbol{\lambda},\boldsymbol{\theta}_{-{s}}}
            p(\mathcal{D}_C \mid H_{1,\theta,{s}})} \\[6pt]
    =&\frac{\max\limits_{\boldsymbol{\lambda},\boldsymbol{\theta}_{-{s}}}
            p(\mathcal{D}_C \mid 
               \boldsymbol{\lambda},\boldsymbol{\theta}_{-{s}},\theta_{{s}}=\theta_s^{hist})}
           {\max\Bigl\{
                \max\limits_{\boldsymbol{\lambda},\boldsymbol{\theta}_{-{s}}}
                   p(\mathcal{D}_C \mid 
                      \boldsymbol{\lambda},\boldsymbol{\theta}_{-{s}},\theta_{{s}}=\theta_s^{hist}+\delta),\;
                \max\limits_{\boldsymbol{\lambda},\boldsymbol{\theta}_{-{s}}}
                   p(\mathcal{D}_C \mid 
                      \boldsymbol{\lambda},\boldsymbol{\theta}_{-{s}},\theta_{{s}}=\theta_s^{hist}-\delta)
              \Bigr\}},
              \nonumber
\end{align*}
where  $\theta_{s}^{hist}$ can be replaced by 
$\hat{\theta}_{{s}}^{hist}=E\{\pi_1(\theta_{s})\}=\mu_{s}$. 

If some PE model time intervals have  few observations,  the LRs for the $\lambda_j$'s may not be robust. To address this, 
we use smoothing to borrow information from neighboring intervals.  For time interval indices $m$ and $j$, 
we use  a discrete triangular distribution function with mode $j$ and non-negative
integer-valued bandwidth $\nu$ as a smoothing kernel,
 (\cite{KOKONENDJI20101655,Wansouwé2025}), 
\begin{align}
    K(m;j,\nu)=\begin{cases} 1-\frac{|j-m|}{\nu+1} & \text{if } \nu\geq 1~\text{and } |m-j| \in \{0,1,\dots,\nu\}\\
    1 & \text{if } \nu=0~ \text{and } m=j\\
     0 & \text{otherwise}.
     \end{cases} \label{eq:kernel1}
\end{align}
The smoothed LR for time interval $I_j$ is
\begin{align}
    \widetilde{LR}_{\lambda,j}(\nu)=\frac{\sum_{m=1}^J r_{m,C}\cdot K(m;j,\nu)\cdot {LR}_{\lambda,m}}{\sum_{m=1}^J r_{m,C}\cdot K(m;j,\nu)}, \label{eq:kernel2}
\end{align}
where $r_{m,C}=\sum_{i:k_i=C}d_{i,m}$ is the  number of events in the $m$th time interval for $C$. 
 Given $\nu$, the closer that $I_m$ and $I_j$  are to each other, and the more events in interval $I_m$, the greater the impact the unsmoothed likelihood ratio ${LR}_{\lambda,m}$ will have in determining $\widetilde{LR}_{\lambda,j}(\nu)$.  We select the bandwidth $\nu$ using leave-one-out cross-validation. Details will be given below, in Section \ref{calibration}. Many other smoothers  incorporate the influence of neighboring intervals in calculating a kernel weight (\cite{KOKONENDJI2011497,parzen1962estimation,Epanechnikov1969,Altman1992,Cleveland1988}).  If desired, a different kernel may be used here. 

The SAM weights are defined using logistic elastic functions (\cite{jiang2023elastic}),  
\begin{equation}
\begin{aligned}
    \widetilde{w}_{\lambda,j}=h_{\lambda,j}(\widetilde{LR}_{\lambda,j}(\nu))=1-\frac{1}{1+\exp\{a_{\lambda,j}+b_{\lambda,j}\log(\widetilde{LR}_{\lambda,j}(\nu))\}},\\
    \widetilde{w}_{\theta,{s}}=h_{\theta,{s}}({LR}_{\theta,{s}})=1-\frac{1}{1+\exp\{a_{\theta,{s}}+b_{\theta,{s}}\log({LR}_{\theta,{s}})\}}, \label{eq:SAMweight}
\end{aligned}
\end{equation}
with tuning parameters $a_{\lambda,j}$, $b_{\lambda,j}$, $a_{\theta,{s}}$ and $b_{\theta,{s}}$  
 selected by simulation-based calibration. Details  are given in Section \ref{calibration}.
The elastic functions $h_{\lambda,j}$ and $h_{\theta,{s}}$  map the prior-data conflict measures $\widetilde{{LR}}_{\lambda,j}$ and $LR_{\theta,{s}}$ to the mixture weights. Each $h\rightarrow 1$ when the LR corresponds to no prior-data conflict, and $h\rightarrow 0$ when the LR corresponds to large prior-data conflict.  

Given  the control arm SAM prior $\pi_{SAM}(\boldsymbol{\lambda},\boldsymbol{\theta})$ and  likelihood  in \eqref{eq:likelihood}, the joint posterior distribution for $(\boldsymbol{\lambda},\boldsymbol{\theta})$ given the trial control arm data and historical control data is
\begin{align}
    p_{SAM}(\boldsymbol{\lambda},\boldsymbol{\theta} \mid   \mathcal{D}_{C},\mathcal{D}^{hist})\propto L(\boldsymbol{\lambda},\boldsymbol{\theta} \mid \mathcal{D}_{C})\ \pi_{SAM}(\boldsymbol{\lambda},\boldsymbol{\theta}).\label{eq:posterior}
\end{align}
Reintroducing conditioning on the terms $\boldsymbol{s}_C$ and $\boldsymbol{s}_E$,  we denote $\boldsymbol{\theta}_{C}=(\theta_{2,C},\dots,\theta_{|\mathcal{G}_{C}|,C})$,  $\boldsymbol{\theta}_{E}=(\theta_{2,E},\dots,\theta_{|\mathcal{G}_{E}|,E})$, $\boldsymbol{\lambda}_C=(\lambda_{1,C},\dots,\lambda_{J,C})$ and $\boldsymbol{\lambda}_E=(\lambda_{1,E},\dots,\lambda_{J,E})$. Since the hazards for  $E$ and $C$  are modeled independently, the same posterior derivation in \eqref{eq:posterior} applies to $E$, with the SAM prior  replaced by a noninformative prior, since  no historical data are available for $E$. 

\subsection{Identifying clusters by Bayesian model averaging}

 We have assumed for simplicity that the cluster assignments 
{$\boldsymbol{s}_C$ and  $\boldsymbol{s}_E$} for the $C$ and $E$ arm hazards  were known. In practice, clusters must be identified adaptively from the data.  To do this, we formulate cluster identification as a Bayesian model averaging (BMA) problem. 
For each arm $k$ = $E$ or $C$, different values of  $\boldsymbol{{s}}_{k}$  create different subgroup clusters, and thus different models. To apply BMA, we  enumerate all possible combinations of $\boldsymbol{{s}}_{k}$, and assign each combination a model index, $\boldsymbol{M}_{1,k},~\boldsymbol{M}_{2,k},\dots,\boldsymbol{M}_{Y_k,k}$, where $Y_k$ denotes the  number of  models determined by  $\boldsymbol{{s}}_{k}$. Since  $\boldsymbol{{s}}_{k}$ corresponds to a partition of the subgroup index set $\{1,\cdots,G\}$,  it follows that 
 $Y_k=B_G$,  the $G^{th}$ Bell number (\cite{Bell1938}). If $G=2$, there are  $B_2=2$ possible partitions, $\{\{1\},\{2\}\}$ and $\{1,2\},$ for each  arm corresponding, respectively, to cluster membership vectors $\boldsymbol{{s}}_k$ =  (1,2) and (1,1), and  induced clusters $\mathcal{G}_k=\{1,2\}$ and $\{1\}$. If $G=3$, there are $B_3=5$ possible partitions, $\{\{1\},\{2\},\{3\}\},~\{\{1,2\},\{3\}\},~\{\{1,3\},\{2\}\},~\{\{2,3\},\{1\}\}$ and $\{1,2,3\}$ for each  arm corresponding, respectively, to cluster membership vectors $\boldsymbol{{s}}_k$ = (1,2,3), (1,1,3), (1,2,1), (1,2,2), and (1,1,1), and induced clusters $\mathcal{G}_k=\{1,2,3\},\{1,3\},\{1,2\},\{1,2\},\{1\}$.

Denote the cluster membership vector for treatment $k$ 
under model $\boldsymbol{M}_{y,k}$ by $\boldsymbol{{s}}^{(y)}_k$. 
We specify a likelihood and prior, and obtain the posteriors $p(\boldsymbol{\lambda}_E,\boldsymbol{\theta}_{E} \mid \mathcal{D}_{E},\boldsymbol{{s}}^{(y)}_E)$  and $p_{SAM}(\boldsymbol{\lambda}_C,\boldsymbol{\theta}_{C}\mid \mathcal{D}_{C},\mathcal{D}^{hist},\boldsymbol{{s}}^{(y)}_C)$ using BMA as follows.
Let ${WAIC}_{y,C}$ denote the  Watanabe–Akaike Information Criterion (\cite{watanabe2010asymptotic}) under model $\boldsymbol{M}_{y,C}$. Given $\boldsymbol{s}^{(y)}_C$ for $y=1,\cdots,Y_C$, 
we define the posterior weights
\begin{align}
\kappa_{y,C}=\frac{\exp(-\frac{1}{2}{WAIC}_{y,C})}{\sum_{l=1}^{Y_C}\exp(-\frac{1}{2}{WAIC}_{l,C})}.\label{eq:BMA-weight}
\end{align}  
This gives the BMA posterior for arm $C$,  
\begin{align}
    \widetilde{p}(\boldsymbol{\lambda}_C,\boldsymbol{\theta}_{C} \mid \mathcal{D}_{C},\mathcal{D}^{hist})=\sum_{y=1}^{Y_C}\kappa_{y,C}\cdot p_{SAM}(\boldsymbol{\lambda}_C,\boldsymbol{\theta}_{C} \mid \mathcal{D}_{C},\mathcal{D}^{hist},\boldsymbol{{s}}^{(y)}_C).
\end{align}
The posterior $\widetilde{p}(\boldsymbol{\lambda}_E,\boldsymbol{\theta}_{E} \mid \mathcal{D}_{E})$ 
  is derived similarly.   
   The weights in \eqref{eq:BMA-weight}  may be defined using other methods, such as the deviance information criterion (DIC) (\cite{spiegelhalter2002bayesian}), or leave-one-out (LOO) information criterion (\cite{vehtari2017practical}).

\section{Trial Design and Conduct}\label{trialdesign}
\subsection{Group Sequential Tests} \label{survival comparison}
 The proposed design's goals are to compare the failure time distributions of $E$ and $C$ for each subgroup, adaptively borrow information from historical control data,  and  cluster subgroups. 
This is done under the failure time model described in Section \ref{model}. 
The design records each patient's subgroup at enrollment, randomizes them between $E$ and $C$, 
and follows the patient to observe their time  to the failure event or censoring, up to a  maximum followup time $T$.
For subgroup $g$, let $N_g$ denote the maximum total sample size, $r_g$ the accrual rate, and $d_g$  the current number of events at an interim decision time, with  $\boldsymbol{d}=(d_1,\dots,d_G)$. At any point during the trial, let $\mathcal{D}_{k,\boldsymbol{d}}$ denote the current survival data for arm $k$, with $\mathcal{D}_{\boldsymbol{d}} = \mathcal{D}_{C,\boldsymbol{d}}\cup\mathcal{D}_{E,\boldsymbol{d}}$ the failure time data of the two arms combined.

The $r^{th}$ interim analysis, for $r=1, \cdots, R$, is done when $d_g$ reaches a prespecified value $d_{r,g}^*$, where $0<d_{1,g}^*<\dots<d_{R,g}^*$, for each undropped subgroup $g\in\{1,\dots,G\}$. Each interim analysis uses the failure time data from all subgroups, with   three possible decisions: 
(1)  conclude  $E$ is superior to $C$
in $g$, and stop enrolling or following patients in $g$;
(2) conclude $E$  is inferior to $C$ in $g$ and stop enrolling or following patients in $g$; or
(3) 
continue enrollment in $g$.
If enrollment to a subgroup is continued beyond the $R^{th}$ interim analysis,  a final $R+1^{st}$ analysis is done for all remaining subgroups when $d^*_{R+1,g}>d^*_{R,g}$ failure events are observed for patients in subgroup $g$. 
The design may adaptively stop enrollment early to subgroups where it has been concluded that one treatment is better than the other, but continue enrollment to the remaining subgroups. This improves  the design's efficiency, and only enrolls patients if  comparisons of $E$ to $C$ in their subgroup have been inconclusive.

At each interim analysis, the design requires two steps to perform each subgroup-specific test. The first step enumerates all possible combinations of the cluster membership vectors
$\boldsymbol{{s}}_C$ and $\boldsymbol{{s}}_E$  based on the current data,  including failure time data of all patients in all subgroups, whether they have been discontinued or not. This is used to identify the models $\boldsymbol{M}_{1,k}, \dots, \boldsymbol{M}_{Y_k,k}$ for $k=C$ and $E$. Under each $\boldsymbol{M}_{y,k}$,  the SAM prior and posterior of $(\boldsymbol{\lambda}_k,\boldsymbol{\theta}_{k})$ are derived, and BMA is used to compute  posteriors.
The second step uses {Restricted~Mean~Survival~Time~(RMST)} (\cite{Royston2013}) to compare the failure time distributions of  $E$ and $C$ in each subgroup. RMST reliably characterizes a between-treatment effect based on failure time data under non-PH, or when the event rate is low (\cite{han2022restricted}). 
To define RMST-based tests, let $\tau$ be a prespecified time horizon of interest for the disease being studied. 

For each arm $k$ and subgroup $g$, let $S_{k,g}(t)$ denote the survival function. Let $h_{k,g}(t)=\lambda_{j,k,g}$ for $t\in I_j$ be the hazard function. Given $\boldsymbol{\lambda}_{k,g}=(\lambda_{1,k,g},\dots,\lambda_{J,k,g})$, the RMST is
\begin{equation}
    \begin{aligned}
        \operatorname{RMST}_{k,g}(\tau)&=\int_0^\tau S_{k,g}(t)dt=\int_0^\tau \exp\{-\sum_{j=1}^J\lambda_{j,k,g}[\min(t,t_j)-t_{j-1}]_+\}dt\\
    &= \sum_{j=1}^{J}
\exp\!\Big(-A_{j-1,k,g}\Big)\,
\frac{1 - \exp\!\big(-\lambda_{j,k,g}\, B_j(\tau)\big)}
{\lambda_{j,k,g}},
    \end{aligned}
\end{equation}
where 
\[
A_{j-1,k,g} = \sum_{m=1}^{j-1} \lambda_{m,k,g} \, (t_m - t_{m-1}), \quad A_{0,k,g}\equiv0,
\quad
\text{and}\quad B_j(\tau) = \big[\min(\tau,t_j) - t_{j-1}\big]_+,
\]
denoting $[x]_+=\max(x,0)$. The RMST difference between $E$ and $C$ for subgroup $g$ is $\psi_g(\tau)$ = $\operatorname{RMST}_{E,g}(\tau)-
\operatorname{RMST}_{C,g}(\tau).$
A large positive value of $\psi_g(\tau)$ implies superiority of $E$ over $C$, and a large negative implies inferiority of $E$. For each subgroup $g$, we perform a one-sided test of the null hypothesis $H_{g,0}:\psi_g(\tau) \leq 0$ 
versus the alternative $H_{g,1}:\psi_g(\tau) > 0$.

Let $c$ be a fixed  cutoff. In each of up to $R$ subgroup-specific interim analyses, there are three  possible  decisions in subgroup $g$.
\begin{enumerate}
    \item  {\it Stop for Superiority}:  If $\Pr\{\psi_g(\tau) > 0 \mid {\cal D}_{\boldsymbol{d}},{\cal D}^{hist}\}>c$  reject $H_{g,0},$ conclude  $E$ is superior to $C$, and stop enrollment to subgroup $g$.
    \item  {\it Stop for Futility}: If $\Pr\{\psi_g(\tau) < 0 \mid {\cal D}_{\boldsymbol{d}},{\cal D}^{hist}\}>c$  accept $H_{g,0}$, conclude $E$ is inferior to $C$, and stop enrollment to subgroup $g$.
    \item  {\it Continue}: If the current data provide insufficient evidence to support a conclusive decision,  continue enrolling patients in $g$.
\end{enumerate}

The trial enrolls patients in any subgroup $g$ that has not been discontinued, up to maximum sample size $N_g$. If at some point there are no active subgroups, the trial is terminated. Otherwise, a final analysis is done when the targeted number of events $d^*_{R+1,g}$ for each undropped subgroup is reached. If a subgroup reaches its target  early, follow-up for this subgroup is stopped and its event time data are not changed thereafter. The procedures for the first and second steps of the final analysis  mirror those of the interim analyses,  with one key difference. At the final analysis, for each undropped subgroup $g$,  if $\Pr\{\psi_g(\tau)>0 \mid {\cal D}_{\boldsymbol{d}},{\cal D}^{hist}\} > c$  $H_{g,0}$ is rejected with the conclusion that $E$ is superior to $C$. Otherwise,  $H_{g,0}$ is accepted. 
 Figure \ref{fig:Fig-1} provides a flowchart illustrating trial conduct.

\subsection{Design Parameter Calibration} \label{calibration}
The design's performance depends on the SAM prior's ability to adaptively borrow external control data, and numerical values of the decision rule cutoff. To obtain good OCs, we use simulation-based cross-validation to select the bandwidth $\nu$ in \eqref{eq:kernel1} and \eqref{eq:kernel2}, and use calibration to establish tuning parameters for the SAM prior and  decision cutoff $c$. 

To determine  $\nu$, we follow \cite{KOKONENDJI2011497} and \cite{Garcia-Portugues2025} by using weighted least square cross-validation to select a value of $\nu$ each time we calculate the smoothed LR in \eqref{eq:kernel1}. Denote the kernel-smoothed likelihood in $I_j$ computed with bandwidth $\nu$ after omitting $I_j$ by 
\begin{align}
    \widetilde{LR}_{\lambda,j,-j}(\nu)=\frac{\sum_{m\neq j} r_{m,C}\cdot K(m;j,\nu)\cdot {LR}_{\lambda,m}}{\sum_{m\neq j} r_{m,C}\cdot K(m;j,\nu)}.
\end{align}
Denoting the cross-validation grid by $\mathcal{H}$,
the selected bandwidth is 
\begin{align}
    \widetilde{\nu}=\arg\min_{\nu\in\mathcal{H}} \sum _{j=1}^J{r_{j,C}\cdot(LR_{\lambda,j}-\widetilde{LR}_{\lambda,j,-j}(\nu))^2}.
\end{align}
Given $\{\widetilde{LR}_{\lambda,j}(\widetilde{\nu})\}$ and $\{LR_{\theta,{\bf s}}\}$, the SAM weights are defined using the logistic elastic functions in \eqref{eq:SAMweight}, with tuning parameters selected using simulation-based calibration to induce strong information borrowing when $\mathcal{D}_C$ and $\mathcal{D}^{hist}$ agree for each parameter, and little information borrowing when they disagree. 

The following steps  determine the parameters in the elastic functions \eqref{eq:SAMweight} needed to compute the weights 
$\widetilde{w}_{\lambda,j}$ of  $\pi_{SAM}(\lambda_j)$ for each  $I_j$. One may generate $\widetilde{w}_{\theta,{s}}$ using the same procedure  used for $\widetilde{w}_{\lambda,j}$.  The following computations, done while planning the trial, 
require a vector $\boldsymbol{d}$ of subgroup-specific event counts to compute the elastic 
function for each interim and final analysis.    
We  set $\boldsymbol{d}$ = $\boldsymbol{d}_r^*=(d_{r,1}^*,\dots,d_{r,G}^*)$ to determine the parameters used in the $r^{th}$ interim analysis, and
 $\boldsymbol{d}$ = $\boldsymbol{d}_{R+1}^*$, the targeted event counts used for the final analysis.    

\begin{enumerate}    
\item {\it Full Agreement Case for Computing Smoothed LRs}. In this case, the trial control arm data ${\cal D}_C$ agrees with ${\cal D}^{hist}$.  Simulate a large number $U$ of replicates of the trial control arm dataset $\mathcal{D}_{C,\boldsymbol{d}}$, which depends on ${\boldsymbol{d}}$,
 when $\lambda_{{k,C}}=\lambda_{k}^{hist}$ for interval $I_k,~k=1,\dots,J$, 
 and $\theta_{{s},C}=\theta_{{s}}^{hist},~{s}=2,\dots,|\mathcal{G}_C|$.  Calculate the smoothed likelihood ratios
$\widetilde{\boldsymbol{{LR}}}_{0,j}=(\widetilde{LR}_{\lambda,j}^{(1)},\dots,\widetilde{LR}_{\lambda,j}^{(U)})$ 
for $I_j$,  where $\widetilde{LR}_{\lambda,j}^{(u)}$ denotes the value of $\widetilde{LR}_{\lambda,j}$ based on the $u$th simulated dataset. For the simulations, the subgroup $g$ sample size is $N_g$ and the enrollment rate is $r_g$.   

\item  {\it Partial Agreement Cases for Computing Smoothed LRs}. In these cases, all parameters except $\lambda_{j,C}$ agree with ${\cal D}^{hist}$. 
Simulate $U$ replicates of $\mathcal{D}_{C,\boldsymbol{d}}$ 
with $\lambda_{{j,C}}=\lambda_{j}^{hist}e^\delta$ and $\lambda_{{k,C}}=\lambda_{k}^{hist},$ for intervals $I_k$,  $~k=1,\dots,j-1,j+1,J$ and $\theta_{{s},C}=\theta_{{s}}^{hist},~{s}=2,\dots,|\mathcal{G}_C|$.  Calculate the LR for $\lambda_{j,C}$, to obtain the smoothed likelihood ratios $\boldsymbol{\widetilde{LR}}_{1,j}^+=(\widetilde{LR}_{\lambda,j}^{(1),+},\dots,\widetilde{LR}_{\lambda,j}^{(U),+})$, where $\widetilde{LR}_{\lambda,j}^{(u),+}$ denotes the value of $\widetilde{LR}_{\lambda,j}$ based on the $u$th simulated $\mathcal{D}_{C,\boldsymbol{d}}$. Repeat this with $\mathcal{D}_{C,\boldsymbol{d}}$ simulated from $\lambda_{{j,C}}=\lambda_{j}^{hist}e^{-\delta}$ while fixing other parameters at their values for the congruent cases, to obtain $\boldsymbol{\widetilde{LR}}_{1,j}^-=(\widetilde{LR}_{\lambda,j}^{(1),-},\dots,\widetilde{LR}_{\lambda,j}^{(U),-})$, where $\widetilde{LR}_{\lambda,j}^{(u),-}$ denotes the value of $\widetilde{LR}_{\lambda,j}$ based on the $u$th simulated $\mathcal{D}_{C,\boldsymbol{d}}$. The value of $\delta$ is the CSD in \eqref{eq:hypothesislambda}.
    \item {\it Elastic Functions for the Likelihood Ratios}. 
  Let $\overline{LR}_{0,j}$ = median($\boldsymbol{\widetilde{LR}}_{0,j}$),   $\overline{LR}_{1,j}^+$  = median($\boldsymbol{\widetilde{LR}}_{1,j}^+$), and 
      $\overline{LR}_{1,j}^-$ = median($\boldsymbol{\widetilde{LR}}_{1,j}^-$),   and $\overline{LR}_{1,j}=\max\{\overline{LR}_{1,j}^+,\overline{LR}_{1,j}^-\}$.
When the trial and historical control data agree  for $\lambda_j$, the SAM weight for 
$\lambda_j$ should be close to 1.   Thus, the LRs in Step 1 above, the full agreement case, 
should be mapped to a SAM weight close to 1.  To do this,  we set the fixed constant  $Q_1$  to a value near 1, such as .9,  and require $h_{\lambda,j}(\overline{LR}_{0,j}) = Q_1$. 
When the trial and historical control data disagree  for $\lambda_j$, the SAM weight for 
$\lambda_j$ should be close to 0. Thus, the LRs in Step 2 above, where there is partial agreement,  should be mapped to a SAM weight close to 0.  We do this by setting the fixed constant $Q_0$ to a value near 0, such as .1, and require $ h_{\lambda,j}(\overline{LR}_{1,j})=Q_0$.
This  determines 
    \begin{align*}
        a_{\lambda,j}&= \frac{\log(\overline{LR}_{1,j}){\operatorname{logit}}(Q_1)-\log(\overline{LR}_{0,j}){\operatorname{logit}}(Q_0)}{\log(\overline{LR}_{1,j}/\overline{LR}_{0,j})} \\
        b_{\lambda,j}&=\frac{{\operatorname{logit}}(Q_0)-{\operatorname{logit}}(Q_1)}{\log(\overline{LR}_{1,j}/\overline{LR}_{0,j})}.
    \end{align*}
\end{enumerate}
Smaller $Q_1$ or smaller $Q_0$ produces more conservative information borrowing. We recommend calibrating interim elastic functions using smaller values, to avoid overly aggressive early borrowing. In our simulation study with one interim analysis, we use $Q_1=.90,~Q_0=.10$ for the final analysis and $Q_1=.80,~Q_0=.0001$ for the interim analysis.

To determine the probability cutoff $c$, we use a simulation-based calibration procedure to obtain a design with good OCs. For each $H_{g,0},~g=1,\dots,G$, the subgroup-specific Type I error rate (SSER) is $\alpha_g=\Pr(\text{reject}~H_{g,0}\mid  H_{g,0})$. Since  $G$ hypotheses are tested simultaneously, we focus on the family-wise Type I error rate (FWER), defined as 
\begin{align}
   FWER   = \Pr(\text{reject at least one}~H_{g,0}\mid\cap_{g=1}^G H_{g,0}).
\end{align}

Under the global null hypothesis, $H_0=\cap_{g=1}^GH_{g,0}$, all subgroups are homogeneous with no treatment effects. 
We specify  subgroup-specific sample sizes, enrollment rates, target event counts at the interim and final analyses, and the time horizon  $\tau$. We then perform a one-dimensional grid search to find the smallest value of $c$ so that FWER $\leq \tilde{\alpha}$ under $H_0$.

\section{Simulation Study}\label{simulationstudy}

\subsection{Design of the Simulations}
We evaluated the OCs of the proposed design, applied to the motivating trial described in Section \ref{motivatingtrial}, indexing the CRT-P and CRT-D subgroups with $g=1$ and $g=2$, respectively, so $G=2$. The individual maximum follow-up time is $T$ = 12 months. 

To derive the informative priors $\pi_1(\boldsymbol{\lambda}_C)$ and $\pi_1(\boldsymbol{\theta}_{C})$ above, we assumed that ${\mathcal{G}}_C$ and $\boldsymbol{s}_C$ were given.  In the simulations, these were determined adaptively.
For $G=2$, there are two possible models for each arm $k=C,E$, either $\boldsymbol{s}_k=(1,1)$ and $\mathcal{G}_k=\{1\}$ where the subgroups have the same arm $k$ distribution with one cluster, or $\boldsymbol{s}_k=(1,2)$ and $\mathcal{G}_k=\{1,2\}$, where the subgroups have different arm $k$ distributions with  two singleton clusters.
For $\boldsymbol{s}_k=(1,1),~\mathcal{G}_k=\{1\}$, the informative prior is  $\pi_1(\boldsymbol{\lambda}_C\mid \boldsymbol{s}_C=(1,1))$.
For $\boldsymbol{s}_k=(1,2),~\mathcal{G}_k=\{1,2\}$, the informative prior is $\pi_1(\boldsymbol{\lambda}_C,\theta_{2,C}\mid \boldsymbol{s}_C=(1,2))=\pi_1(\boldsymbol{\lambda}_C\mid \boldsymbol{s}_C=(1,2))\times \pi_1(\theta_{2,C})$. Estimated survival distributions are available for the CRT-P subgroup ($g=1$) from the MIRACLE trial (\cite{Abraham2002}), whereas the CRT-D subgroup ($g=2$)’s historical study, MIRACLE-ICD (\cite{young2003}), does not report them. Therefore, for $\boldsymbol{s}_C=(1,1)$,  
 we applied  empirical Bayes by exploiting gamma-exponential conjugacy to obtain 
$$
\pi_1(\boldsymbol{\lambda}_C \mid \boldsymbol{s}_C=(1,1))=\prod_{j=1}^4\pi_1(\lambda_{j,C}\mid \boldsymbol{s}_C=(1,1))
$$
for time intervals  $I_1=[0,2),~I_2=[2,6),~I_3=[6,12),~I_4=[12,\infty)$ with $\pi_1(\lambda_{1,C}\mid\boldsymbol{s}_C=(1,1))$ = $\operatorname{Gamma}$$(20.01,426.27),$ $\pi_1(\lambda_{2,C}\mid\boldsymbol{s}_C=(1,1))$ = $\operatorname{Gamma}$(24.01, 693.96), and $\pi_1(\lambda_{3,C}\mid\boldsymbol{s}_C=(1,1))=\pi_1(\lambda_{4,C}\mid\boldsymbol{s}_C=(1,1)) = 
\operatorname{Gamma}(24.01,964.22)$. The interval partition was informed by the temporal hazard patterns described in the motivating trial (\cite{Curtis2013}), and was further slightly refined to ensure approximately balanced event counts across intervals for stable estimation under the piecewise exponential model.
For $\boldsymbol{s}_C=(1,2)$,  $\pi_1(\boldsymbol{\lambda}_C\mid\boldsymbol{s}_C=(1,2))=\pi_1(\boldsymbol{\lambda}_C\mid\boldsymbol{s}_C=(1,1))$ with $\pi_1(\theta_{2,C})$  =  
$\mathcal{N}(0,\sigma^2)$ with  $\sigma^2=10000$,
since there is no historical information for the between-subgroup effect for $C$. 
The historical mean estimates 
are $\hat{\lambda}_{1}^{hist}=0.047,~\hat{\lambda}_{2}^{hist}=0.035,~\hat{\lambda}_{3}^{hist}=0.025,~\hat{\lambda}_{4}^{hist}=0.025$.

In our simulations, we compared $\text{SAM}_\text{BMA}$ to several competing methods,
all using the same group sequential  procedure. 
The first comparator,  $\text{NP}_\text{HETERO}$, uses a noninformative prior, assumes heterogeneous subgroups, and does not borrow external control information. The second  comparator,  $\text{NP}_\text{HOMO}$, 
uses a noninformative prior,  assumes 
 homogeneous subgroups and does not borrow external control information. The third comparator,  $\text{PP}_\text{HOMO}$, 
uses a power prior to borrow external control information while assuming 
   homogeneous subgroups, with the power parameter assigned a $Unif(0,1)$ prior. As a benchmark, we included the $\text{NP}_\text{BMA}$ design, which uses the same BMA procedure as $\text{SAM}_\text{BMA}$
 but does not borrow external control information. Table \ref{table:illustration} illustrates these approaches in terms of how they borrow information from external controls and between subgroups.

In the simulations, we assumed the subgroups have prevalence 0.50 each.  
We evaluated  all designs under the 16 scenarios given in Table {\ref{table:scenarios}}.  
We calibrated $c$ to control  the one-sided FWER  $\tilde{\alpha} \leq 0.05$,  in scenario 1, with one interim analysis for each subgroup, when $d_{1,1}=\lceil d^*_{2,1}/2\rceil$ and $d_{1,2}=\lceil d^*_{2,2}/2\rceil$ respectively. We assumed accrual rates $r_1=15/\text{month}$, $r_2=15/\text{month}$, $d^*_{2,1}=120$, $d^*_{2,2}=120$, $N_1=400$, and $N_2=400$.

We generated event times assuming a PH model with a PE hazard,
\begin{align}
    h_{k,g}(t)={\lambda}_{j,k
}\exp(\theta_{g,k}),~t\in I_j,~j=1,\dots,4,~g=1,2,~k=C,E,\label{eq:simulationPE}
\end{align}
where $t$ is time in months and $\theta_{1,k}=0$ with subgroup 1   the baseline. We set
   $\lambda_{j,C}=\hat{\lambda}_{j}^{hist}\cdot\exp(\Delta_j)$ 
   for $j=1,\dots,4,$
where $\Delta_j$ quantifies prior-data conflict for interval $I_j$ in the control arm. We fixed the informative priors while varying  $(\Delta_1,\Delta_2,\Delta_3)$ to represent different levels of 
trial-versus-external control
data conflict. Since $I_4=[12,\infty)$ is beyond the maximum follow-up time $T=12$, $\Delta_4$ does not affect the results, and we fixed $\Delta_4=0$. We varied the values of $\boldsymbol{\lambda}_{E}$, $\theta_{C,2}$ and $\theta_{E,2}$, which represent different levels of subgroup heterogeneity and $E$ versus $C$ effects. 
In all scenarios, non-informative priors $\pi_0(\lambda_j)=\operatorname{Gamma}(0.01,0.01)$ and $\pi_0(\theta_{{s}})=\mathcal{N}(0,10000)$ were assumed, and we used $\delta=0.45$ as the value for a {{clinically significant difference}}. Under each scenario, 1,000 trials were simulated for each design.

\subsection{Simulation Results} \label{simresult}
Tables \ref{table:result1-8}, \ref{table:result9-16}, along with Tables~S1 and~S2 in the Supplementary Material, report the OCs of the designs under the 16 scenarios given in Table \ref{table:scenarios}. 
Overall, $\text{SAM}_\text{BMA}$ does a good job of adaptively clustering subgroups when they are homogeneous, and  adaptively borrowing external control information when there is little prior-data conflict between the trial and external controls.  $\text{SAM}_\text{BMA}$  also has robust performance when $h_{C,g}(t)$ and $h_{E,g}(t)$ are non-proportional. 

In scenario 1, the null case with PH, and scenario 2, the null case where 
$h_{E,g}(t)$ and $h_{C,g}(t)$ are non-proportional,
all methods controlled the FWER near .05. 
In scenarios 3-8, where the external control dataset is congruent with trial controls and there are $E$-versus-$C$ effects,   $\text{SAM}_\text{BMA}$ has much higher SSP than $\text{NP}_\text{HETERO}$ and $\text{NP}_\text{BMA}$  for subgroups where $E$ is superior to $C$, and larger correct early termination rates.  
In scenario 3, the two subgroups have the same control hazard functions,  they are congruent with the historical data, and $E$ is superior to $C$ in both subgroups. In this scenario, it is desirable to combine the data of the two subgroups within each arm to improve efficiency,  and borrow external control information. The SSPs of $\text{NP}_\text{HOMO}$, $\text{PP}_\text{HOMO}$, $\text{NP}_\text{BMA}$ and $\text{SAM}_\text{BMA}$  for $g=1,2$ all outperform  $\text{NP}_\text{HETERO}$  by at least 18\%. Borrowing from the external control data further enhances the SSPs, with power 0.89 in $g=1$ of $\text{SAM}_\text{BMA}$   compared to 0.80 for $\text{NP}_\text{BMA}$. 

In scenarios 4-7, $E$ is superior to $C$ in one subgroup but has no effect or is inferior to $C$ in the other.  In these scenarios, 
$\text{SAM}_\text{BMA}$ has much higher SSP than all other designs, and
$\text{NP}_\text{HOMO}$ and $\text{PP}_\text{HOMO}$ both perform poorly because they assume homogeneity. 
In scenario 8, where $E$ is superior to $C$ in subgroup 1 but inferior to $C$ in subgroup $2$,
a qualitative treatment-subgroup interaction, 
 $\text{SAM}_\text{BMA}$ has by far the largest SSP in subgroup 1.

In scenarios 9-16, where the external control data are incongruent with the trial control data,  $\text{SAM}_\text{BMA}$ has better type I error control and greater SSP  compared with 
$\text{PP}_\text{HOMO}$. In scenarios 9-10, where subgroups are homogeneous and there is  no $E$-versus-$C$ treatment effect, $\text{SAM}_\text{BMA}$ outperforms $\text{PP}_\text{HOMO}$ in controlling the FWER. In scenarios 11-12, where the subgroups are homogeneous and $E$ is superior to $C$, $\text{SAM}_\text{BMA}$  has SSPs  similar to $\text{NP}_\text{BMA}$.
$\text{PP}_\text{HOMO}$  has slightly higher SSPs than $\text{SAM}_\text{BMA}$. This is because the posterior distributions $\psi_g(\tau),~g=1,2$ have smaller variances due to  ignoring the subgroups. In
these scenarios,  the homogeneity assumption of $\text{PP}_\text{HOMO}$  happens to be correct, but in any case with heterogenous effects across subgroups this assumption is wrong and $\text{PP}_\text{HOMO}$ is very likely to make incorrect decisions.

In scenarios 13-15, the trial and external controls are incongruent, and $E$ is superior to $C$ in one subgroup but has no effect or is inferior in the other.
Although $\text{SAM}_\text{BMA}$ adaptively downweights the informative component in the mixture prior, it  still has a risk of incorrectly borrowing from the  external control data. The effects of this are small, however.
In scenario 13, the SSP of subgroup 1 is 0.778 for $\text{SAM}_\text{BMA}$ compared with 0.801 for $\text{NP}_\text{BMA}$.  In scenario 14, the SSER of subgroup 1 is inflated to 0.102 for $\text{SAM}_\text{BMA}$ compared to 0.043 for $\text{NP}_\text{BMA}$. In scenario 15, the PETF of subgroup 1 decreases to 0.453 for $\text{SAM}_\text{BMA}$ compared with 0.510 for  $\text{NP}_\text{BMA}$. 
Thus, $\text{SAM}_\text{BMA}$ pays a small price for its risk of incorrectly borrowing from the external controls when there is incongruence.

To assess the sensitivity of the design to the specification of a clinically significant difference, we considered $\delta=0.4$ and $\delta=0.5$. The results,  shown in the bar plots in {Figure S1} in the Supplementary Material, are similar to those for $\delta=0.45$. 

\section{Discussion}
\label{s:discuss}
We have proposed a Bayesian group sequential hybrid control design for trials with time-to-event outcomes that accounts for possible heterogeneity of treatment effects on survival time between subgroups. The design  dynamically borrows external control data using a  self-adapting mixture prior, with  the external data prior given a large weight when the trial-versus-external data conflict is small, and a small weight when the conflict is large. The design thus addresses uncertainty regarding whether the outcome distributions of treatments $E$ or $C$ differ among specified subgroups, as well as the degree to which the trial and external controls agree. For each decision, the design adaptively borrows information across subgroups in each arm using a latent subgroup variable and Bayesian model averaging. It makes subgroup-specific survival comparisons based on the posterior distributions of restricted mean survival time, and includes adaptive subgroup-specific interim stopping if superiority or inferiority of $E$ compared to $C$ is concluded in that subgroup. 

Our computer simulations  show that, compared to competing methods, under a wide range of scenarios the proposed design  does a good job of adaptively borrowing information between subgroups and from external control data. It reliably borrows information between subgroups when homogeneity is found to be likely, while reliably separating subgroups when heterogeneity is detected. In many scenarios, it has substantially higher subgroup-specific power than competing designs that account for heterogeneity but do not assume a self-adapting mixture prior.  
The design  also adapts well to trial-versus-external data disagreement, improving power and early stopping rates when the disagreement is small, and controlling type I error inflation and power loss when the prior-data conflict is large.

The effective sample size (ESS) provides a useful tool to quantify the amount of information borrowed from historical data, yet its formulation for mixture priors is nontrivial. \cite{wiesenfarth2020quantification} demonstrated that the approach of \cite{morita2008determining} does not adequately capture the information structure of mixture priors. Methods proposed by \cite{Schmidli2014} and \cite{gravestock2019power} are more appropriate for mixtures but lack data adaptivity, whereas the data-dependent measures of \cite{wiesenfarth2020quantification} and \cite{reimherr2021prior} better reflect the adaptive borrowing mechanism underlying SAM priors. However, ESS is inherently sensitive to the choice of reference noninformative prior and the information metric, and no consensus currently exists. Developing coherent and interpretable ESS measures for mixture priors therefore remains an unresolved issue and a meaningful direction for future research.

Our proposed design accounts for subgroups but does not incorporate additional covariates, 
which might improve efficiency of information borrowing.  Following \cite{wang2019propensity}, \cite{wang2022propensity}, and \cite{zhao2025ps}, this may be done  using propensity score matching to identify a subset of external control patients  comparable to those in the current control arm with respect to the covariates. A  SAM 
prior then might be formulated to borrow information from this matched subset for decision making in the hybrid control design.

\section*{Data availability statement}
Data sharing is not applicable in this paper as all data in this paper are computer simulated.

\begin{spacing}{1.0}
\bibliography{citations_v11}
\end{spacing}

\clearpage
\begin{table}[htbp]
\centering
\caption{Competing designs studied in the simulations. NP  = non-informative prior, 
PP = power prior, SAM = self-adapting mixture prior, BMA = Bayesian model averaging, HOMO = assume homgeneous subgroups,
HETERO = assume heterogeneous subgroups. }
 \label{table:illustration}
\begin{tabular}{ccc}
\toprule
\textbf{Design} & \textbf{Subgroup Assumptions} & \textbf{Use External Control Data?}\\
\midrule
$\text{NP}_\text{HOMO}$ & Ignore Subgroups & No\\
$\text{NP}_\text{HETERO}$ & Account for Subgroups & No\\
$\text{PP}_\text{HOMO}$ & Ignore Subgroups & Yes \\
$\text{NP}_\text{BMA}$ & Subgroup Clustering via BMA & No\\
$\text{SAM}_\text{BMA}$ & Subgroup Clustering via BMA & Yes\\

\bottomrule
\end{tabular}
\end{table}
\clearpage

\begin{table}[htbp]
\centering
\caption{Simulation scenarios with clinically significant difference $\delta=0.45$. $\Delta_j$   
= $\log(\lambda_{j,C}/\hat{\lambda}_{j}^{hist}),~j=1,2,3$; $\lambda_{j,C}=$ hazard  for $g=1$ in $I_j$ for  arm $C$; $\theta_{2,C}=$ log hazard ratio of $g=2$ vs. $g=1$ for arm $C$; $\lambda_{j,E}=$ hazard  of $g=1$ for arm $E$; $\theta_{2,E}=$ log hazard ratio of $g=2$ vs. $g=1$ for arm $E$; $\psi_g(\tau)=$ restricted mean survival time (RMST) difference between  $E$ and $C$ in subgroup $g$ = 1 or 2.}
 \label{table:scenarios}
\small
\begin{tabular}{ccccccc}
\toprule
\textbf{Scen} & $(\boldsymbol{\Delta}_1,\boldsymbol{\Delta}_2,\boldsymbol{\Delta}_3)$ &$(\boldsymbol{\lambda_{1,C},\lambda_{2,C},\lambda_{3,C}})$& $\boldsymbol{\theta_{2,C}}$ &$(\boldsymbol{\lambda_{1,E},\lambda_{2,E},\lambda_{3,E}})$ & $\boldsymbol{\theta_{2,E}}$ & $\boldsymbol{\psi}_g(\tau)$\\
\midrule
\multicolumn{5}{l}{\textbf{\underline{Trial and External Controls Are Congruent}}} \\
1 & (0.00, 0.00, 0.00) & (.047,.035,.025) & 0.00 & (.047,.035,.025) & 0.00 & (0.00,0.00)  \\ 
        2 & (0.05, 0.00, -0.05) & (.049,.035,.024) & 0.00 & (.060,.031,.015) & 0.00 & (0.00,0.00)  \\ 
        3 & (0.00, -0.03, -0.05) & (.047,.034,.024) & 0.00 & (.028,.020,.014) & 0.00 & (0.81, 0.81)  \\ 
        4 & (0.05, 0.00, -0.05) & (.049,.035,.024) & 0.00 & (.028,.020,.014) & 0.55 & (0.90, 0.00)  \\ 
        5 & (0.00, -0.03, -0.05) & (.047,.034,.024) & -0.55 & (.033,.018,.009) & 0.00 & (0.87, 0.00)  \\ 
        6 & (0.00, 0.00, 0.00) & (.047,.035,.025) & 0.00 & (.057,.031,.017) & -0.55 & (0.00, 0.89)  \\ 
        7 & (0.00, 0.00, 0.00) & (.047,.035,.025) & 0.00 & (.070,.052,.037) & -0.90 & (-0.92, 0.82)  \\ 
        8 & (-0.05, -0.03, 0.00) & (.045,.034,.025) & -0.70 & (.027,.020,.015) & 0.70 & (0.80, -1.46)  \\ 
\hline
\\
\multicolumn{5}{l}{\textbf{\underline{Trial and External Controls Are Incongruent}}} \\
9 & (-0.45, -0.45, -0.45) & (.030,.022,.016) & 0.00 & (.030,.022,.016) & 0.00 & (0.00,0.00)  \\ 
        10 & (-0.50, -0.45, -0.40) & (.028,.022,.017) & 0.00 & (.023,.022,.024) & 0.00 & (0.00, 0.00)  \\ 
        11 & (0.55, 0.50, 0.45) & (.081,.057,.039) & 0.00 & (.045,.038,.032) & 0.00 & (1.09, 1.09)  \\ 
        12 & (0.60, 0.45, 0.00) & (.085,.054,.025) & 0.00 & (.057,.035,.015) & 0.00 & (1.02, 1.02)  \\ 
        13 & (0.60, 0.45, 0.40) & (.085,.054,.037) & -0.51 & (.049,.033,.024) & 0.00 & (1.22, 0.00)  \\ 
        14 & (-0.65, -0.55, -0.45) & (.024,.020,.016) & 0.65 & (.030,.019,.011) & 0.00 & (0.00, 1.07)  \\ 
        15 & (0.00, -0.45, -0.60) & (.047,.022,.014) & 0.00 & (.081,.031,.014) & -1.05 & (-0.84, 0.79)  \\ 
        16 & (-0.65, -0.65, -0.45) & (.024,.018,.016) & 0.75 & (.042,.038,.034) & -0.60 & (-1.09, 1.15)  \\ 
\bottomrule
\end{tabular}
\end{table}

\clearpage
\begin{table}[!htbp]
\caption{Operating characteristics of the  five designs under congruent scenarios 1-8  in Table \ref{table:scenarios}.   NP = non-informative prior, PP = power prior, Dur = mean trial duration. For subgroup $g$ = 1 or 2, $\text{SSP}_g$ = Pr(reject $H_{0,g}$), FWER = family-wise error rate, $\text{MSS}_g$ = mean sample size.}
 \label{table:result1-8}
 \small
 \centering
    \begin{tabular}{cccccccc}
    \toprule
        \textbf{Scen} & \textbf{Method} &{$\textbf{SSP}_1$} & {$\textbf{SSP}_2$} & \textbf{FWER} & {$\textbf{MSS}_1$}&{$\textbf{MSS}_2$}& \textbf{Dur} \\ 
        \midrule 
        1 & $\text{NP}_\text{HOMO}$ & 0.051 & 0.051 & 0.051 & 392 & 392 & 34.5  \\ 
        ~ & $\text{PP}_\text{HOMO}$ & 0.051 & 0.051 & 0.051 & 393 & 393 & 34.5  \\ 
        ~ & $\text{SAM}_\text{BMA}$ & 0.027 & 0.028 & 0.050 & 396 & 395 & 35.2  \\ 
        ~ & $\text{NP}_\text{BMA}$ & 0.029 & 0.027 & 0.050 & 396 & 395 & 35.2  \\ 
        ~ & $\text{NP}_\text{HETERO}$ & 0.027 & 0.024 & 0.051 & 395 & 395 & 35.2  \\ 
        2 & $\text{NP}_\text{HOMO}$ & 0.051 & 0.051 & 0.051 & 393 & 393 & 36.9  \\ 
        ~ & $\text{PP}_\text{HOMO}$ & 0.048 & 0.048 & 0.048 & 393 & 393 & 36.9  \\ 
        ~ & $\text{SAM}_\text{BMA}$ & 0.029 & 0.027 & 0.049 & 396 & 396 & 37.6  \\ 
        ~ & $\text{NP}_\text{BMA}$ & 0.031 & 0.026 & 0.053 & 396 & 396 & 37.7  \\ 
        ~ & $\text{NP}_\text{HETERO}$ & 0.027 & 0.023 & 0.050 & 395 & 395 & 37.7  \\ 
        3 & $\text{NP}_\text{HOMO}$ & 0.927 & 0.927 & 0.927 & 338 & 338 & 25.5  \\ 
        ~ & $\text{PP}_\text{HOMO}$ & 0.975 & 0.975 & 0.975 & 329 & 329 & 23.6  \\ 
        ~ & $\text{SAM}_\text{BMA}$ & 0.889 & 0.806 & 0.953 & 340 & 348 & 29.4  \\ 
        ~ & $\text{NP}_\text{BMA}$ & 0.795 & 0.775 & 0.907 & 353 & 353 & 31.9  \\ 
        ~ & $\text{NP}_\text{HETERO}$ & 0.612 & 0.617 & 0.849 & 366 & 368 & 36.4  \\ 
        4 & $\text{NP}_\text{HOMO}$ & 0.474 & 0.474 & 0.474 & 370 & 370 & 33.7  \\ 
        ~ & $\text{PP}_\text{HOMO}$ & 0.538 & 0.538 & 0.538 & 367 & 366 & 32.7  \\ 
        ~ & $\text{SAM}_\text{BMA}$ & 0.826 & 0.065 & 0.828 & 348 & 393 & 34.9  \\ 
        ~ & $\text{NP}_\text{BMA}$ & 0.745 & 0.063 & 0.752 & 357 & 393 & 35.8  \\ 
        ~ & $\text{NP}_\text{HETERO}$ & 0.693 & 0.025 & 0.703 & 356 & 396 & 36.2  \\ 
        5 & $\text{NP}_\text{HOMO}$ & 0.521 & 0.521 & 0.521 & 393 & 393 & 34.6  \\ 
        ~ & $\text{PP}_\text{HOMO}$ & 0.768 & 0.768 & 0.768 & 387 & 387 & 31.2  \\ 
        ~ & $\text{SAM}_\text{BMA}$ & 0.874 & 0.050 & 0.877 & 385 & 399 & 39.0  \\ 
        ~ & $\text{NP}_\text{BMA}$ & 0.747 & 0.056 & 0.756 & 390 & 399 & 39.2  \\ 
        ~ & $\text{NP}_\text{HETERO}$ & 0.673 & 0.026 & 0.684 & 389 & 399 & 39.5  \\ 
        6 & $\text{NP}_\text{HOMO}$ & 0.467 & 0.467 & 0.467 & 373 & 373 & 34.1  \\ 
        ~ & $\text{PP}_\text{HOMO}$ & 0.545 & 0.545 & 0.545 & 367 & 367 & 32.7  \\ 
        ~ & $\text{SAM}_\text{BMA}$ & 0.067 & 0.754 & 0.763 & 395 & 355 & 36.8  \\ 
        ~ & $\text{NP}_\text{BMA}$ & 0.062 & 0.721 & 0.732 & 395 & 359 & 36.9  \\ 
        ~ & $\text{NP}_\text{HETERO}$ & 0.023 & 0.684 & 0.690 & 397 & 358 & 37.4  \\ 
        7 & $\text{NP}_\text{HOMO}$ & 0.050 & 0.050 & 0.050 & 378 & 395 & 38.3  \\ 
        ~ & $\text{PP}_\text{HOMO}$ & 0.052 & 0.052 & 0.052 & 379 & 396 & 38.5  \\ 
        ~ & $\text{SAM}_\text{BMA}$ & 0.000 & 0.770 & 0.770 & 342 & 352 & 31.2  \\ 
        ~ & $\text{NP}_\text{BMA}$ & 0.000 & 0.733 & 0.733 & 350 & 356 & 32.3  \\ 
        ~ & $\text{NP}_\text{HETERO}$ & 0.000 & 0.619 & 0.619 & 354 & 362 & 33.7  \\ 
        8 & $\text{NP}_\text{HOMO}$ & 0.000 & 0.000 & 0.000 & 383 & 383 & 36.0  \\ 
        ~ & $\text{PP}_\text{HOMO}$ & 0.007 & 0.007 & 0.007 & 393 & 393 & 38.2  \\ 
        ~ & $\text{SAM}_\text{BMA}$ & 0.713 & 0.000 & 0.713 & 360 & 331 & 31.9  \\ 
        ~ & $\text{NP}_\text{BMA}$ & 0.595 & 0.000 & 0.595 & 372 & 332 & 34.2  \\ 
        ~ & $\text{NP}_\text{HETERO}$ & 0.614 & 0.000 & 0.614 & 366 & 325 & 32.8  \\ 
        \bottomrule
    \end{tabular}
\end{table}

\appendix

\begin{table}[!ht]
\caption{Operating characteristics of the  five designs under incongruent scenarios 9-16  in Table \ref{table:scenarios}.   NP = non-informative prior, PP = power prior, Dur = mean trial duration. For subgroup $g$ = 1 or 2, $\text{SSP}_g$ = Pr(reject $H_{0,g}$), FWER = familywise error rate, $\text{MSS}_g$ = mean sample size.}
 \label{table:result9-16}
 \small
 \centering
    \begin{tabular}{cccccccc}
    \toprule
        \textbf{Scen} & \textbf{Method} &{$\textbf{SSP}_1$} & {$\textbf{SSP}_2$} & \textbf{FWER} & {$\textbf{MSS}_1$}&{$\textbf{MSS}_2$}& \textbf{Dur} \\ 
        \midrule 
        9 & $\text{NP}_\text{HOMO}$ & 0.041 & 0.041 & 0.041 & 398 & 398 & 38.8  \\ 
        ~ & $\text{PP}_\text{HOMO}$ & 0.158 & 0.158 & 0.158 & 395 & 395 & 37.8  \\ 
        ~ & $\text{SAM}_\text{BMA}$ & 0.068 & 0.044 & 0.092 & 397 & 398 & 39.4  \\ 
        ~ & $\text{NP}_\text{BMA}$ & 0.026 & 0.026 & 0.047 & 399 & 399 & 39.6  \\ 
        ~ & $\text{NP}_\text{HETERO}$ & 0.025 & 0.028 & 0.053 & 399 & 399 & 39.8  \\ 
        10 & $\text{NP}_\text{HOMO}$ & 0.048 & 0.048 & 0.048 & 397 & 397 & 38.9  \\ 
        ~ & $\text{PP}_\text{HOMO}$ & 0.166 & 0.166 & 0.166 & 394 & 394 & 37.6  \\ 
        ~ & $\text{SAM}_\text{BMA}$ & 0.082 & 0.035 & 0.100 & 396 & 398 & 39.5  \\ 
        ~ & $\text{NP}_\text{BMA}$ & 0.024 & 0.026 & 0.046 & 399 & 398 & 39.7  \\ 
        ~ & $\text{NP}_\text{HETERO}$ & 0.024 & 0.029 & 0.053 & 398 & 398 & 39.7  \\ 
        11 & $\text{NP}_\text{HOMO}$ & 0.941 & 0.941 & 0.941 & 264 & 264 & 18.1  \\ 
        ~ & $\text{PP}_\text{HOMO}$ & 0.841 & 0.841 & 0.841 & 297 & 297 & 20.6  \\ 
        ~ & $\text{SAM}_\text{BMA}$ & 0.807 & 0.793 & 0.914 & 295 & 290 & 21.7  \\ 
        ~ & $\text{NP}_\text{BMA}$ & 0.796 & 0.775 & 0.904 & 293 & 293 & 21.8  \\ 
        ~ & $\text{NP}_\text{HETERO}$ & 0.637 & 0.625 & 0.862 & 316 & 311 & 23.8  \\ 
        12 & $\text{NP}_\text{HOMO}$ & 0.930 & 0.930 & 0.930 & 283 & 283 & 19.9  \\ 
        ~ & $\text{PP}_\text{HOMO}$ & 0.830 & 0.830 & 0.830 & 312 & 312 & 22.3  \\ 
        ~ & $\text{SAM}_\text{BMA}$ & 0.778 & 0.765 & 0.903 & 316 & 315 & 24.3  \\ 
        ~ & $\text{NP}_\text{BMA}$ & 0.768 & 0.753 & 0.897 & 316 & 316 & 24.3  \\ 
        ~ & $\text{NP}_\text{HETERO}$ & 0.606 & 0.596 & 0.835 & 339 & 337 & 26.7  \\ 
        13 & $\text{NP}_\text{HOMO}$ & 0.565 & 0.565 & 0.565 & 334 & 349 & 28.0  \\ 
        ~ & $\text{PP}_\text{HOMO}$ & 0.431 & 0.431 & 0.431 & 349 & 367 & 30.0  \\ 
        ~ & $\text{SAM}_\text{BMA}$ & 0.778 & 0.057 & 0.785 & 325 & 392 & 33.4  \\ 
        ~ & $\text{NP}_\text{BMA}$ & 0.801 & 0.050 & 0.806 & 316 & 393 & 33.4  \\ 
        ~ & $\text{NP}_\text{HETERO}$ & 0.758 & 0.025 & 0.767 & 317 & 395 & 33.7  \\ 
        14 & $\text{NP}_\text{HOMO}$ & 0.681 & 0.681 & 0.681 & 388 & 388 & 32.3  \\ 
        ~ & $\text{PP}_\text{HOMO}$ & 0.855 & 0.855 & 0.855 & 383 & 383 & 29.6  \\ 
        ~ & $\text{SAM}_\text{BMA}$ & 0.102 & 0.904 & 0.910 & 396 & 382 & 38.3  \\ 
        ~ & $\text{NP}_\text{BMA}$ & 0.043 & 0.886 & 0.891 & 399 & 383 & 39.2  \\ 
        ~ & $\text{NP}_\text{HETERO}$ & 0.026 & 0.832 & 0.840 & 399 & 383 & 39.4  \\ 
        15 & $\text{NP}_\text{HOMO}$ & 0.030 & 0.030 & 0.030 & 399 & 399 & 39.2  \\ 
        ~ & $\text{PP}_\text{HOMO}$ & 0.078 & 0.078 & 0.078 & 399 & 399 & 38.8  \\ 
        ~ & $\text{SAM}_\text{BMA}$ & 0.000 & 0.780 & 0.780 & 394 & 389 & 36.0  \\ 
        ~ & $\text{NP}_\text{BMA}$ & 0.000 & 0.751 & 0.751 & 393 & 390 & 36.0  \\ 
        ~ & $\text{NP}_\text{HETERO}$ & 0.000 & 0.647 & 0.647 & 393 & 391 & 37.0  \\ 
        16 & $\text{NP}_\text{HOMO}$ & 0.064 & 0.064 & 0.064 & 396 & 396 & 38.6  \\ 
        ~ & $\text{PP}_\text{HOMO}$ & 0.129 & 0.129 & 0.129 & 393 & 393 & 37.9  \\ 
        ~ & $\text{SAM}_\text{BMA}$ & 0.000 & 0.875 & 0.875 & 354 & 342 & 32.9  \\ 
        ~ & $\text{NP}_\text{BMA}$ & 0.000 & 0.868 & 0.868 & 347 & 344 & 32.3  \\ 
        ~ & $\text{NP}_\text{HETERO}$ & 0.000 & 0.879 & 0.879 & 339 & 335 & 30.9  \\  
        \bottomrule
    \end{tabular}
    
\end{table}

\clearpage

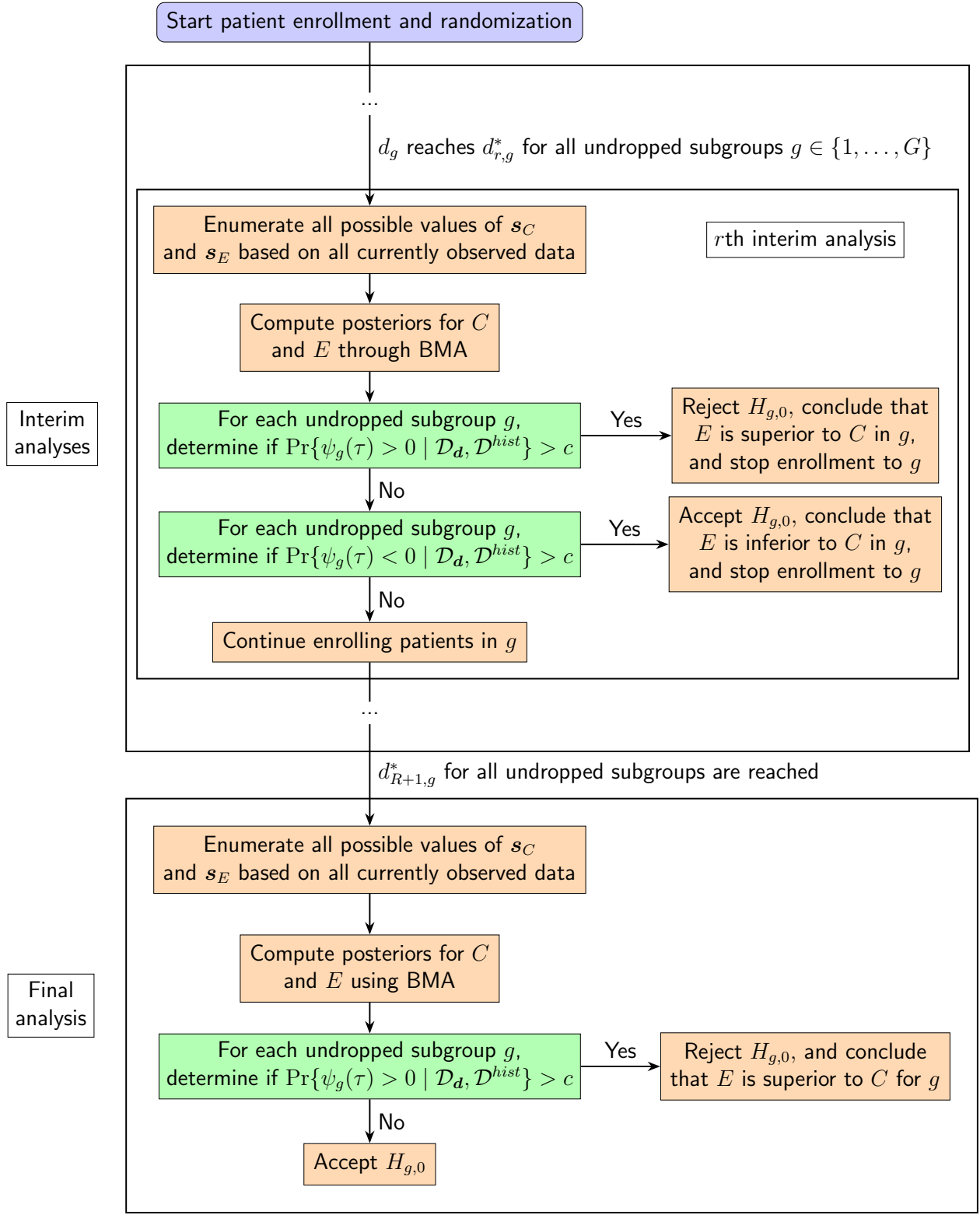
\begin{figure}[ht]
    \resizebox{\textwidth}{!}{\begin{tikzpicture}[scale=1]
    \node(starttrial) [startstop] {Start patient enrollment and randomization};
    \node(help1)[annotate1,below of =starttrial,yshift=-0.5cm]{...};
    \node(group)[process,below of = help1,yshift=-1.5cm]{Enumerate all possible values of $\boldsymbol{{s}}_C$\\and $\boldsymbol{{s}}_E$ based on all currently observed data};
    \node(average)[process,below of = group,yshift=-0.8cm]{Compute posteriors for $C$\\and $E$ through BMA};
    \node(pro1)[process1,below of =average,yshift=-0.8cm]{For each undropped subgroup $g$,\\determine if $\Pr\{\psi_g(\tau)>0 \mid {\cal D}_{\boldsymbol{d}},{\cal D}^{hist}\} > c$};
    \node(sup)[process,right of =pro1,xshift=7cm]{Reject $H_{g,0},$ conclude that\\$E$ is superior to $C$ in $g$,\\and stop enrollment to $g$};
    \node(pro2)[process1,below of =pro1,yshift=-1cm]{For each undropped subgroup $g$,\\determine if $\Pr\{\psi_g(\tau)<0 \mid {\cal D}_{\boldsymbol{d}},{\cal D}^{hist}\} > c$};
    \node(fut)[process,right of =pro2,xshift=7cm]{Accept $H_{g,0},$ conclude that\\$E$ is inferior to $C$ in $g$,\\and stop enrollment to $g$};
    \node(cont)[process,below of =pro2,yshift=-0.8cm]{Continue enrolling patients in $g$};
    \node(help2)[annotate1,below of =cont,yshift=-0.3cm]{...};
    \node(groupFA)[process,below of = help2,yshift=-1.7cm]{Enumerate all possible values of $\boldsymbol{{s}}_C$\\and $\boldsymbol{{s}}_E$ based on all currently observed data};
    \node(averageFA)[process,below of = groupFA,yshift=-1cm]{Compute posteriors for $C$\\and $E$ using BMA};
    \node(proFA)[process1,below of =averageFA,yshift=-0.8cm]{For each undropped subgroup $g$,\\determine if $\Pr\{\psi_g(\tau)>0 \mid {\cal D}_{\boldsymbol{d}},{\cal D}^{hist}\} > c$};
    \node(supFA)[process,right of =proFA,xshift=7cm]{Reject $H_{g,0},$ and conclude\\that $E$ is superior to $C$ for $g$};
    \node(acceptFA)[process,below of =proFA,yshift=-0.8cm]{Accept $H_{g,0}$};
    \node[draw=black, thick, inner sep=0.5cm, 
     fit=(groupFA)(averageFA) (proFA)(supFA)(acceptFA), 
      name=analysisboxFA] {};
    \node[draw=black, thick, inner sep=0.3cm, 
      fit=(group) (average) (sup) (fut)(pro1)(pro2)(cont) , 
      name=analysisboxIA] {};
    \node[draw=black, thick, inner sep=0.5cm, 
      fit=(group) (sup) (fut)(pro1)(pro2)(cont)(help1)(help2) , 
      name=IA] {};
    \node(IAi)[annotate,right of =group,xshift=7cm]{$r$th interim analysis};
    \node(IA)[annotate,left of =analysisboxIA,xshift=-8.1cm]{Interim\\analyses};
    \node(FA)[annotate,left of =analysisboxFA,xshift=-8.2cm]{Final\\analysis};

      \draw[thick](starttrial)--(help1);
      \draw [arrow] (help1) -- node [anchor=west,yshift=0.2cm]{$d_g$ reaches $d_{r,g}^*$ for all undropped subgroups $g\in\{1,\dots,G\}$}(group); 
      \draw[arrow] (group) -- (average);
      \draw[arrow](average)--(pro1);
      \draw[arrow] (pro1) --node[anchor=south]{Yes}(sup);
      \draw[arrow] (pro1) --node[anchor=west]{No} (pro2);
      \draw[arrow](pro2)--node[anchor=south]{Yes}(fut);
      \draw[arrow](pro2)--node[anchor=west]{No}(cont);
      \draw[thick](cont)--(help2);
      \draw[arrow](help2) -- node[anchor=west]{$d^*_{R+1,g}$ for all undropped subgroups are reached}(groupFA);
      \draw[arrow](groupFA)--(averageFA);
      \draw[arrow](averageFA)--(proFA);
      \draw[arrow](proFA)--node[anchor=south]{Yes}(supFA);
      \draw[arrow](proFA)--node[anchor=west]{No}(acceptFA);
\end{tikzpicture}}
    \caption{Schematic of trial conduct}
    \label{fig:Fig-1}
\end{figure}

\end{document}